\documentclass[a4paper,DIV12]{scrartcl}

\usepackage{amsmath,graphicx,url,booktabs}

\newcommand*{\name}[1]{\textsc{#1}}                 

\DeclareMathOperator{\Li}{Li}                       
\DeclareMathOperator{\BrAv}{BrAv}                   
\DeclareMathOperator{\sign}{sign}                   

\newcommand*{\I}{i}                                 
\newcommand*{\SP}[2]{\ensuremath{#1{\cdot}#2}}      
\newcommand*{\hc}{\mathrm{h.\,c.}}                  
\newcommand*{\abs}[1]{\left|#1\right|}              
\newcommand*{\conv}{\ast}                           
\renewcommand*{\Re}{\mathrm{Re}}                    

\newcommand*{\LQCD}%
            {\ensuremath{\Lambda_{\text{QCD}}}}     
\newcommand*{\Heff}%
            {\ensuremath{\mathcal{H}_\text{eff}}}   
\newcommand*{\alphas}%
            {\ensuremath{\alpha_{\mathrm{s}}}}      
\newcommand*{\alphaEM}{\ensuremath{\alpha_{\mathrm{em}}}} 
\newcommand*{\GF}{\ensuremath{G_{\mathrm{F}}}}      
\newcommand*{\CF}{\ensuremath{C_{\mathrm{F}}}}      
\newcommand*{\Nc}{\ensuremath{N_{\mathrm{c}}}}      

\newcommand*{\EW}{\mathrm{EW}}
\newcommand*{\CP}{\mathrm{CP}}

\newcommand*{\Ai}[1][]{A^{\mathrm{i}#1}}
\newcommand*{\Af}[1][]{A^{\mathrm{f}#1}}

\newcommand*{\Amp}{\ensuremath{\mathcal{A}}}

\newcommand*{\ptcl}[1]{\ensuremath{{#1}}}
\newcommand*{\B}{\ptcl{B}}
\newcommand*{\Bbar}{\ensuremath{\bar{\B}}}
\newcommand*{\Bm}{\ensuremath{\ptcl{B}^-}}
\newcommand*{\Bdq}{\ensuremath{\Bbar^0}}
\newcommand*{\Bsq}{\ensuremath{\Bbar_s}}
\newcommand*{\K}{\ptcl{K}}
\newcommand*{\Kbar}{\ensuremath{\bar{\K}}}
\newcommand*{\prho}{\ptcl{\rho}}
\newcommand*{\pomega}{\ptcl{\omega}}
\newcommand*{\pphi}{\ptcl{\phi}}
\newcommand*{\ppi}{\ptcl{\pi}}

\renewcommand*{\d}{d}                   
\newcommand*{\e}{\mathrm{e}}            
\newcommand*{\FMslash}[1]{\not\!#1}

\begin{document}
\renewcommand{\topfraction}{1}


\begin{titlepage}

  \begin{flushright}
    PITHA~06/13\\
    hep-ph/0612290\\[0.1cm]
    December 21, 2006
  \end{flushright}

  \vspace{0.7cm}
  \begin{center}
    \Large\bfseries\boldmath
    Branching fractions, polarisation and asymmetries\\ of  
    $B\to VV$ decays
    \unboldmath
  \end{center}

  \vspace{1.5\baselineskip}
  \begin{center}
    \name{Martin Beneke}$^a$,
    \name{Johannes Rohrer}$^a$ and
    \name{Deshan Yang}$^b$\footnote{Address after December~1:
      Department of Physics, Graduate School of the Chinese Academy of
      Sciences, Beijing 100049, P.\,R.\,China}
    \\[1.5\baselineskip]
    \itshape
    ${}^a$Institut f\"ur Theoretische Physik E, RWTH Aachen,\\
    D--52056 Aachen, Germany\\[\baselineskip]
    ${}^b$Theoretical Particle Physics Group (Eken), Nagoya University,\\
    Furo-cho, Chikusa-ku, Nagoya 464-8602, Japan
  \end{center}

  \vspace{2\baselineskip}
  \begin{abstract}
  \noindent
  We calculate the hard-scattering kernels relevant to the
  negative-helicity decay amplitude in $\B$ decays to two vector
  mesons in the framework of QCD factorisation. We then perform a
  comprehensive analysis of the 34 $B\to VV$ decays, including $B_s$
  decays and the complete set of polarisation observables. We find
  considerable uncertainties from weak annihilation and the
  non-factorisation of spectator-scattering. Large longitudinal
  polarisation is expected with certainty only for a few
  tree-dominated colour-allowed modes, which receive small penguin and
  spectator-scattering contributions.  This allows for an accurate
  determination of the CKM angle $\alpha$ (or $\gamma$) from
  $S_{L}^{\rho\rho}$ resulting in $\alpha=(85.6^{+7.4}_{-7.3})^\circ$
  We also emphasize that the $\rho K^*$ system is ideal for an
  investigation of electroweak penguin effects.
  \end{abstract}

  \vfill
\end{titlepage}


\setcounter{page}{1}

\section{Introduction}

The variety of accessible final states in $B$ decays to two mesons
provides an abundant source of information on CP violation and
flavour-changing processes. When the final state consists of two
vector mesons, an angular analysis of the vector mesons' decay
products also provides insight into the spin structure of the
flavour-changing interaction. For the $V-A$ coupling of the Standard
Model, a specific pattern of the three helicity amplitudes is
expected~\cite{Koerner1979}, such that the longitudinal polarisation
fraction $f_L$ should be close to 1. Since $f_L\approx 0.5$ was first
observed \cite{Aubert:2003mm,Chen:2003jf} for penguin-dominated
strangeness-changing decays, many theoretical papers addressed the
question whether this result could be explained as a
strong-interaction effect, or whether it could be reproduced within
specific ``New Physics'' scenarios
\cite{Cheng:2001aa,Li:2003he,Kagan:2004uw,Colangelo:2004rd,Hou:2004vj,Ladisa:2004bp,Rohrer:DT,Li:2004ti,Yang:2004pm,Das:2004hq,Li:2004mp,Kim:2004wq,Zou:2005gw,Huang:2005if,Baek:2005jk,Huang:2005qb,BRYEWP,Chen:2006jz,Chang:2006dh}.

In this paper we revisit this question using the QCD factorisation
framework \cite{BBNS1999,BBNS2000} to deal with the strong interaction
in the amplitude calculation.  Our study goes beyond previous ones in
several respects. On the theoretical side we provide the first
complete results for the hard-scattering kernels relevant to
vector-vector ($VV$) final states, correcting several errors in the
literature. (In fact, the only correct calculation
is~\cite{Kagan:2004uw}.) We also provide a more detailed discussion of
the factorisation structure and power counting for the various
amplitudes. It seems to have escaped attention so far that, contrary
to the longitudinal polarisation amplitude and those relevant to $PP$
and $PV$ final states, the transverse polarisation amplitudes do not
factorise even at leading power in the heavy-quark expansion. This,
together with the high sensitvity to penguin
weak-annihilation~\cite{Kagan:2004uw}, implies that the calculation of
polarisation observables stands on a much less solid footing than the
calculation of $\B\to PP, PV$ decays. On the phenomenological side, we
provide estimates for all $\B\to VV$ decays (including $\B_s$ decays)
and for all parameters that enter the angular analysis. Previous
studies concentrated on single or a few decay modes and considered the
longitudinal polarisation fraction $f_L$ only, making it often
difficult to distinguish general patterns from the consequences of
particular parameters choices.

The organisation of the paper is as follows: In Section~\ref{sec:defs} 
we summarise the definitions for the helicity amplitudes, angular variables 
and polarisation observables. The calculation of the 
$\B\to VV$ decay amplitudes in the QCD factorisation framework 
is briefly reviewed in Section~\ref{sec:factorisation}. We then 
discuss a few aspects of the transverse polarisation amplitudes 
that allow for an understanding of the main characteristics of 
$\B\to VV$ phenomenology. One important conclusion from this 
discussion is that the analysis of $B\to VV$ decays will be much 
less rigorous and much more uncertain 
than the corresponding analysis of $B\to PP$ and 
$B\to VV$ modes \cite{BN2003}. The technical results of the 
calculation are summarised in an Appendix. Section~\ref{sec:inputs} 
provides the list of input parameters, an overview of the flavour 
amplitude parameters with theoretical uncertainties, and a
classification of the 34 $B\to VV$ decay channels, which guides 
the subsequent numerical analysis. We begin the analysis 
in Section~\ref{sec:tree} with a discussion of branching fractions, 
CP asymmetries and polarisation observables of 
the nine tree-dominated decays. Among these 
the four colour-allowed modes can be well predicted. In particular, 
we show that the time-dependent CP asymmetry measurement in 
$B^0\to\rho^+\rho^-$ leads to one of the most accurate determinations 
of the CKM angle $\gamma$. In Section~\ref{sec:penguin} we turn 
to the 14 colour-allowed penguin-dominated decay modes. It will be 
seen that theoretical calculations allow for large transverse 
polarisation within large uncertainties. This suggests to determine 
the transverse penguin amplitude from data using the well-measured 
$\phi K^*$ modes. This approach is used to sharpen the 
predictions for the remaining decay modes in this class. The 
analysis concludes in Section~\ref{sec:others} with a brief discussion
of the remaining penguin-dominated modes, and decays 
that occur only through weak annihilation. 
Section~\ref{sec:conclusion} summarises our main results 
and conclusions.

\section{Helicity amplitudes and polarisation observables}
\label{sec:defs}

We consider a $\B$ meson with four-momentum $p_B$ and mass $m_B$
decaying into two light vector mesons $V_1(p_1,\eta^*)$, $V_2
(p_2,\epsilon^*)$ with masses $m_{1,2}$ of order $\LQCD$.  The decay
amplitude can be decomposed into three scalar amplitudes $S_{1,2,3}$
according to
\begin{equation}
  \Amp_{\B\to V_1 V_2} 
    = \I \eta^{*\mu} \epsilon^{*\nu}\left(
      S_1\, g_{\mu\nu} - S_2\, \frac{p_{B\mu}p_{B\nu}}{m_B^2}
      + S_3\, \I\varepsilon_{\mu\nu\rho\sigma} 
      \frac{p_1^\rho p_2^\sigma}{\SP{p_1}{p_2}}
      \right).
  \label{eq:lorentzdecomp}
\end{equation}
with convention $\varepsilon_{0123}=1.$ 
Alternatively, one can choose a basis of amplitudes describing
decays to final state particles with definite helicity
\begin{equation}
  \begin{aligned}
    \Amp_0   &=  \Amp(\B\to V_1(p_1,\eta_0^*)V_2(p_2,\epsilon_0^*))
             = \frac{i m_B^2}{2m_1 m_2}
                 \left(S_1-\frac{S_2}{2}\right)\\
    \Amp_\pm &=  \Amp(\B\to V_1(p_1,\eta_\pm^*)V_2(p_2,\epsilon_\pm^*))
             = i\, (S_1 \mp S_3).
  \end{aligned}
\end{equation}
or use the transversity amplitudes, where $\Amp_{\pm}$ are replaced by
$\Amp_{\parallel} = (\Amp_+ +\Amp_-)/\sqrt{2}$ and $\Amp_{\perp} =
(\Amp_+ - \Amp_-)/\sqrt{2}$, corresponding to linearly polarised final
states. We choose $\vec{p}_2$ to be directed in positive $z$-direction
in the $\B$ meson rest frame, and the polarisation four-vectors of the
light vector mesons such that in a frame where both light mesons have 
large momentum along the $z$-axis, they are given by $\epsilon_\pm^\mu = 
\eta_\mp^\mu = (0,\pm 1,\I,0)/\sqrt{2}$, and 
$\epsilon_0^\mu = p_2^\mu/m_2$, $\eta_0^\mu = p_1^\mu/m_1$. 
Here and thoughout the paper we neglect corrections 
$m_{1,2}^2/m_B^2$ quadratic in the light meson masses. Thus 
$p_{1,2}^\mu = m_B n_\mp^\mu/2$ with
$n_\pm^\mu=(1,0,0,\pm 1)$. Our conventions imply the 
identity
\begin{equation}
  i \varepsilon_{\mu\nu\rho\sigma} \epsilon_\pm^{*\rho} n_+^\sigma 
  = (\mp 1)\,\left(
      n_{+\mu} \epsilon_{\pm \nu}^* -n_{+\nu} \epsilon_{\pm \mu}^*
    \right).
\end{equation} 

\begin{figure}
  \centering
  \includegraphics[width=.55\linewidth]{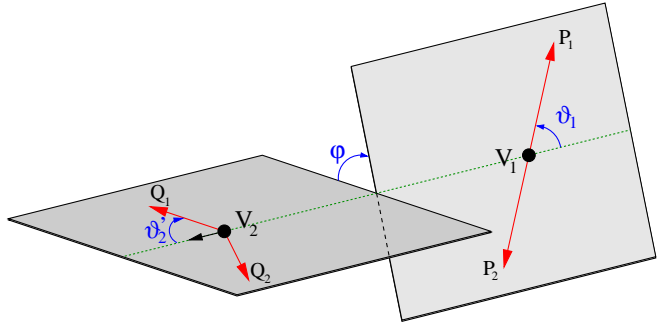}\hfill
  \caption{Decay kinematics in the rest frame of $V_1$.}
  \label{fig:kinematics}
\end{figure}

Experimentally, the magnitudes and relative phases of the
various amplitudes are extracted from the angular distributions of the vector
resonance decay products. The full angular dependence of the cascade
where both vector mesons decay into pseudoscalar particles is given by
\cite{Koerner1979}
\begin{equation}
  \begin{split}
    \frac{\d{\Gamma_{\B\to V_1 V_2 \to\ldots}}}%
         {\d{\cos\vartheta_1}\d{\cos\vartheta_2}\d{\varphi}}
       &\,\propto \,\abs{\Amp_0}^2\cos^2\vartheta_1\cos^2\vartheta_2
             + \frac{1}{4}\sin^2\vartheta_1\sin^2\vartheta_2
               \left( \abs{\Amp_+}^2 + \abs{\Amp_-}^2 \right)
               \\&\mathrel{\hphantom{\propto}}
             - \cos\vartheta_1\sin\vartheta_1\cos\vartheta_2\sin\vartheta_2
               \left[  \Re\left(\e^{-i\varphi} \Amp_0 \Amp_+^*\right)
                     + \Re\left(\e^{+i\varphi} \Amp_0 \Amp_-^*\right)
               \right]
               \\&\mathrel{\hphantom{\propto}}
             + \frac{1}{2} \sin^2\vartheta_1\sin^2\vartheta_2
               \,\Re\left(\e^{2 i\varphi} \Amp_+ \Amp_-^*\right),
  \end{split}
\label{angulardist}
\end{equation}
where $\varphi$ measures the angle between the decay planes of the two
vector mesons in the $\B$ meson rest frame, 
and $\vartheta_{1,2}$ are the angles between the direction
of motion of one of the $V_{1,2}\to PP$ pseudoscalar final states and
the inverse direction of motion of the $\B$ meson as measured in the
$V_{1,2}$ rest frame, see Figure~\ref{fig:kinematics}.
The omitted proportionality factor is such that 
one obtains the decay rate $\Gamma(\B\to V_1 V_2)$ 
when $\cos\vartheta_1$ and 
$\cos\vartheta_2$ are integrated from $-1$ to $1$, and 
$\varphi$ from 0 to $2\pi$.

Thus, any given $\B\to VV$ decay allows us to define five observables
corresponding to the three magnitudes and two relative phases of the
helicity amplitudes, or the five angular coefficients in
(\ref{angulardist}). In experimental analyses, observables are
preferably defined in terms of the transversity amplitudes as they
have definite CP transformation properties. A typical set of
observables consists of the branching fraction, two out of the three
polarisation fractions $f_L, f_\parallel, f_\perp$, and two phases
$\phi_\parallel$, $\phi_\perp$, where
\begin{equation}
  \label{eq:polobs}
  f^{\B}_{L,\parallel,\perp} = \frac{\abs{\Amp_{0,\parallel,\perp}}^2}{
  \abs{\Amp_0}^2+\abs{\Amp_\parallel}^2+\abs{\Amp_\perp}^2},
  \qquad\qquad
  \phi^{\B}_{\parallel,\perp} = \arg\frac{\Amp_{\parallel,\perp}}{\Amp_0}.
\end{equation}
It is conventional to combine the five observables of some $\B\to VV$
decay with those of its CP-conjugate $\Bbar$ decay, and to quote the
ten resulting observables as CP-averages and CP-asymmetries.  We
denote $\Bbar$ decay helicity amplitudes as $\bar{\Amp}_h$ and define
the corresponding transversity amplitudes as
$\bar{\Amp}_{\parallel/\perp} = (\bar{\Amp}_- \pm
\bar{\Amp}_+)/\sqrt{2}$, so that
$\Amp_{\parallel/\perp}=\bar{\Amp}_{\parallel/\perp}$ in the absence
of CP violation. Observables $f_h^{\Bbar}, \phi^{\Bbar}_h$ are then
defined as in (\ref{eq:polobs}), and CP averages and asymmetries
are calculated by
\begin{equation}
    f_h = \frac{1}{2} \left( f_h^{\Bbar}+f_h^{\B} \right),
    \qquad
    A_{\CP}^{h} = \frac{f_h^{\Bbar}-f_h^{\B}}{f_h^{\Bbar}+f_h^{\B}}
\end{equation}
($h=L,\parallel,\perp$) for the polarisation fractions and
\begin{equation}
  \begin{aligned}
    \phi_h &\equiv \phi^{\Bbar}_h - \Delta\phi_h \pmod{2\pi} \\
           &\equiv \phi^{\B}_h + \Delta\phi_h \pmod{2\pi},
     \qquad  -\frac{\pi}{2} \leq \Delta\phi_h < \frac{\pi}{2}
  \end{aligned}
\label{phasedef}
\end{equation}
($h=\,\parallel,\perp$) for the phase observables $\phi_h$ and
$\Delta\phi_h$. The implicit definition (\ref{phasedef}) ensures 
that the CP-averaged phase is the geometrical bisection of the 
\emph{acute} angle enclosed by $\phi^{\B}$ and $\phi^{\Bbar}$; 
the magnitude of this angle is $2\abs{\Delta\phi}$. More explicitly, 
the averaged quantities can be obtained as
  \begin{equation}
    \begin{aligned}
      \phi_h &=
         \frac{1}{2}\left(\phi^{\Bbar}_h+\phi^{\B}_h\right)
        -\pi\cdot\sign\left(\phi^{\Bbar}_h+\phi^{\B}_h\right)
         \theta\left(\abs{\phi^{\Bbar}_h-\phi^{\B}_h}-\pi\right),\\
      \Delta \phi_h &=
         \frac{1}{2}\left(\phi^{\Bbar}_h-\phi^{\B}_h\right)
        +\pi\cdot\theta\left(\abs{\phi^{\Bbar}_h-\phi^{\B}_h}-\pi\right)
        \pmod{2\pi}.\\
    \end{aligned}
  \end{equation}
Our phase convention for the amplitudes and definition of
  observables is compatible with that used in the relevant
  publications of the \name{BaBar} and \name{Belle} collaborations
  (for example
  in~\cite{Aubert:2004xc,Gritsan:2004vs,Chen:2005zv,Abe:2004ku}),
  except for the sign of $\Amp_0$ relative to the transverse
  amplitudes, which leads to an offset of $\pi$ for
  $\phi_{\parallel/\perp}$. We favour the above convention, 
  because it implies $\phi_\parallel=\phi_\perp=0$ and
  $\Delta\phi_\parallel=\Delta\phi_\perp=0$ at leading order, 
  where all strong phases are zero.

\section{\boldmath$\B\to VV$ amplitudes}
\label{sec:factorisation}

The decay amplitudes follow from the matrix elements 
$\langle V_1 V_2|\Heff|\Bbar\rangle$ of the effective 
Hamiltonian (conventions as in \cite{BBNS2001})
\begin{equation}
  \label{eq:Heff}
  \Heff =
     \frac{\GF}{\sqrt{2}} \sum_{p=u,c} \lambda^{(D)}_p
     \left\{
        C_{1} Q_{1}^p + C_{2} Q_{2}^p +\!\!
        \sum_{i=3,\ldots 10,7\gamma,8g}\!\! C_i Q^p_i
     \right\} + \hc
\end{equation}
with $D \in \{d,s\}$ and $\lambda^{(D)}_p = V_{pb}V^*_{pD}$.
A quark model~\cite{Koerner1979} or naive factorisation analysis
indicates a hierarchy of helicity amplitudes
\begin{equation}
  \label{eq:hierarchy}
  \bar{\Amp}_0 : \bar{\Amp}_- : \bar{\Amp}_+ = 
  1 : \frac{\LQCD}{m_b} : \left(\frac{\LQCD}{m_b}\right)^{\!2}
\end{equation}
for $\Bbar$ meson decays. (For $\B$ decays exchange 
$-\leftrightarrow +$.) This is a consequence of the left-handedness 
of the weak interaction and the fact that high-energy QCD interactions 
conserve helicity.

In naive factorisation one considers only the four-quark operators in 
$\Heff$ and approximates their matrix elements by the matrix elements 
of two currents \cite{Fakirov:1977ta}. The helicity 
amplitudes $\Amp_{\Bbar\to V_1 V_2}^h$ are proportional to  
\begin{equation}
  A^h_{V_1 V_2} \equiv 
        \frac{\GF}{\sqrt{2}} \;
        \langle V^h_1|(\bar{q}_s b)_{V-A}|\Bbar_{q_s} \rangle
        \langle V^h_2|(\bar{q} q')_{V}|0\rangle
\end{equation}
in this approximation. Evaluating this expression (conventions for 
the form factors as in \cite{BF2000}) we obtain 
\begin{align}
  \label{eq:Aexpl}
  A^0_{V_1 V_2}   &= \frac{i \GF}{\sqrt{2}}
                     m_B^2 f_{V_2} A_0^{\B\to V_1}(0),
  \qquad 
  A^\pm_{V_1 V_2} = \frac{i \GF}{\sqrt{2}} m_B m_2 f_{V_2}
                     F_{\pm}^{\B\to V_1}(0)
\end{align}
with the definitions 
\begin{equation}
   F_{\pm}^{\B\to V_1}(q^2)
     \equiv (1+\frac{m_1}{m_B}) A_1^{\B\to V_1}(q^2) 
          \mp (1-\frac{m_1}{m_B}) V^{\B\to V_1}(q^2).
  \label{eq:Fpmdef}
\end{equation}
The transverse amplitudes $A^\pm_{V_1 V_2}$ are suppressed by a factor
$m_2/m_B$ relative to $A^0_{V_1 V_2}$. In addition, the axial-vector
and vector contributions to $F_+^{\B\to V_1}(0)$ cancel in the
heavy-quark limit, due to an exact form factor relation
\cite{BF2000,Burdman:2000ku}.  Thus $F_-/A_0 \sim 1$, $F_+/A_0\sim
\mathcal{O}(\LQCD/m_B)$, and (\ref{eq:hierarchy}) follows.

The dominance of the longitudinal amplitude indicated
by~(\ref{eq:hierarchy}) leads to the well-known expectation that $f_L$
should be close to unity. Experimental data for penguin-dominated $\B$
decays is in conflict with this expectation thus motivating
theoretical studies beyond the naive-factorisation approximation.

\subsection{The QCD factorisation approach for $\B\to VV$}

We use the QCD factorisation approach \cite{BBNS1999,BBNS2000} 
to compute the matrix elements $\langle V_1V_2|Q_i|\Bbar\rangle$ of 
the effective Hamiltonian. In this framework they can be expressed 
(at leading power in an expansion of the amplitude in 
$\LQCD/m_B$)  in terms of form factors, meson
light-cone distribution amplitudes and perturbatively calculable hard
scattering kernels. In condensed notation, the factorisation formula
reads
\begin{equation}
  \label{eq:QCDFformula}
  \begin{split}
    \langle V_1 V_2|Q_i|\Bbar\rangle = 
      &\left( F^{\B\to V_1} \,T^I_{i} \conv f_{V_2} \Phi_{V_2} 
               + [V_1\leftrightarrow V_2] \right) 
      + T^{II}_i \conv f_B \Phi_{\B} 
                 \conv f_{V_1}\Phi_{V_1}
                 \conv f_{V_2}\Phi_{V_2},
  \end{split}
\end{equation}
where the star products imply an integration over light-cone
momentum fractions. In addition the framework contains estimates 
of some power corrections, which usually cannot be computed 
rigorously. 

We follow closely the scheme developed in \cite{BN2003} for 
$B\to PP, PV$ decays to match contributions to the hard-scattering
kernels $T^{I,II}_i$ on terms involving products of flavour coefficients
$\alpha_i^{p,h}(V_1 V_2)$ and factorised matrix elements 
$A^h_{V_1 V_2}$. The longitudinal amplitude $h=0$ can be deduced 
from the results given in \cite{BN2003}. For the analysis of the 
present paper we calculated the transverse helicity amplitudes. 
In the following we describe the basic results and main differences 
with respect to the longitudinal amplitude; the expressions for 
the hard-scattering functions are given in the Appendix. 

Non-leptonic decay amplitudes are sums of products of CKM factors, 
Wilson coefficients from (\ref{eq:Heff}) and  
matrix elements (\ref{eq:QCDFformula}) of operators with different 
flavours. It is convenient to organise the amplitudes according 
to flavour. Thus, one writes for example,  
\begin{eqnarray}
   \sqrt2\,{\cal A}_{\B^-\to\rho^0 K^{*-}}^h
   &=& \sum_{p=u,c} \lambda_p^{(s)} 
   \bigg\{A^h_{\rho\bar K^*} \left[ \delta_{pu}\,(\alpha_1^h+\beta_2^h) 
    + \alpha_4^{p,h} + \alpha_{4,{\rm EW}}^{p,h} + \beta_3^{p,h}
    + \beta_{3,{\rm EW}}^{p,h} \right] \nonumber\\
   &&+\, A^h_{\bar K^*\rho} \Big[ \delta_{pu}\,\alpha_2^h 
    + \frac{3}{2}\alpha_{3,{\rm EW}}^{p,h} \Big]\bigg\}.
\label{ampexample}
\end{eqnarray}
In naive factorisation, the flavour coefficients
$\alpha_i^{p,h}(V_1 V_2)$ are linear combinations of Wilson coefficients 
$C_i$. In QCD factorisation, they include non-factorisable 
loop effects and spectator-scattering. The $\beta_i^{p,h}$ coefficients 
parameterise weak annihilation amplitudes. The decomposition of the 
amplitudes for the 34 $VV$ final states in terms of
these quantities follows from the $PV$ expressions given
in~\cite{BN2003} with obvious replacements of pseudoscalar by vector 
mesons. The $\alpha_i$ relate to the coefficients $a_i$ used in the 
older factorisation literature as follows (helicity indices 
and $V_1 V_2$ arguments suppressed):
\begin{equation}
  \label{eq:alphaa}
  \begin{split}
    \alpha_1          &= a_1 \\
    \alpha_2          &= a_2
  \end{split}
  \qquad
  \begin{split}
    \alpha^p_3        &= a^p_3    +              a^p_5 \\
    \alpha^p_{3,\EW}  &= a^p_9    +              a^p_7
  \end{split}
  \qquad
  \begin{split}
    \alpha^p_4        &= a^p_4    - r_\chi^{V_2} a^p_6 \\
    \alpha^p_{4,\EW}  &= a^p_{10} - r_\chi^{V_2} a^p_8,
  \end{split}
\end{equation}
where we have used the notation
\begin{equation}
\label{rchidef}
  r_\chi^V \equiv \frac{2m_V}{m_b} \frac{f_V^\perp}{f_V}.
\end{equation}

The explicit expressions for the negative-helicity coefficients
$a_i^{p-}$ and the transverse weak annihilation amplitudes are
collected in the Appendix. Beyond leading order the $a_i^{p-}$ are
sums of vertex corrections, penguin contractions, and
spectator-scattering contributions, see (\ref{ai}).  Most of the
relevant hard-scattering functions have been calculated before. In
\cite{Cheng:2001aa,Li:2003he,Yang:2004pm,Das:2004hq,Zou:2005gw}
results for all three of these contributions have been given.  We do
not find agreement with these results, however. As far as we
understand, the origin of the discrepancy is that the authors of these
papers use an incorrect projection on the light-cone distribution
amplitudes of transversely polarised vector mesons, which neglects the
transverse-momentum derivative terms in (\ref{eq:Mperp}). An exception
is \cite{Das:2004hq}, which does state the correct projector, but the
results still differ from ours, particularly for the
spectator-scattering contributions.  Kagan \cite{Kagan:2004uw} has
calculated the QCD penguin contractions, spectator-scattering terms 
as well as the weak annihilation amplitudes, but did not consider 
the vertex contractions.  We confirm his results on the penguin 
contractions and QCD penguin annihilation, for which explicit 
expressions were given in the paper.

\subsection{Anatomy of transverse amplitudes}
\label{subsec:transverseAnatomy}

The NLO calculation of the negative-helicity amplitude is quite
similar to the calculation of the longitudinal amplitude. The result
exhibits, however, some qualitative differences which have important
consequences for the phenomenology of $\B\to VV$ decays. In this
section we explain the non-factorisation of the negative-helicity
amplitude; that the positive-helicity amplitude cannot be calculated
in an analogous manner; that the amplitude hierarchy
(\ref{eq:hierarchy}) is violated by electromagnetic effects; that
penguin annihilation is comparatively more significant for transverse
polarisation than longitudinal polarisation penguin amplitudes.

\subsubsection{Non-factorisation of spectator scattering}
\label{sec:spectscat}

The factorisation formula (\ref{eq:QCDFformula}) contains two
structurally different terms, the first of which is dominated by soft
interactions within the $\B\to M_1$ transitions. These are absorbed
into the QCD form factor.  The second term stands for interactions
where a hard (more precisely, hard-collinear) interaction with the
spectator quark in the $\B$ meson takes place. Both terms are of the
same order in the heavy-quark expansion. This remains true for the
transverse polarisation amplitudes, but now one finds that the
convolution integrals over the light-cone distribution amplitudes are
logarithmically divergent due to the occurence of the integral
\begin{equation}
\int_0^1 \d x\,\frac{\phi_1^\perp(x)}{(1-x)^2},
\end{equation} 
see~(\ref{eq:Hm1}),~(\ref{eq:Hm2}). This endpoint divergence at $x=1$
signals that the presumed factorisation of spectator-scattering does
not hold {\em even at leading power}.  A similar effect occurs in the
$\B\to PP, PV$ and longitudinal $\B\to VV$ amplitude only when one
attempts to calculate power corrections by applying the light-cone
projection including twist-3 terms. It is perhaps not surprising that
this divergence is obtained at leading-power for the transverse
amplitude, since the entire amplitude is formally a twist-3 term.
(This is the origin of the power suppression of the transverse
amplitudes relative to the longitudinal amplitude.)
Factorisation-violation at the leading power implies that the
calculation of transverse polarisation amplitudes is on a much less
solid footing than of the other amplitudes, and often should be
considered more as an estimate. In practice, we find that the
non-factorisation of spectator-scattering is only significant for the
colour-suppressed tree and the flavour-singlet QCD penguin amplitudes,
where the (regulated) divergent integral is multiplied by a large
Wilson coefficient. The fact that there are endpoint divergences in
the spectator-scattering contribution to the transverse helicity
amplitudes has been observed in previous calculations, but its
significance for the theoretical status of the factorisation approach
and its phenomenological implications have not been sufficiently
emphasized.

\subsubsection{The positive-helicity amplitude}
\label{subsubsec:positiveHelicity}

The calculation of the kernels $T^I_i$ in (\ref{eq:QCDFformula}) 
can be interpreted as matching the operators $Q_i$ to 
four-quark operators with field content $[\bar\chi\chi] 
[\bar\xi h_v]$ in soft-collinear effective theory 
\cite{Chay:2003ju}. 
The field $\chi$ describes collinear quarks moving in the 
direction of $V_2$, the meson that does not pick up the spectator quark, 
and satisfies $\FMslash{n}_+\chi=0$. The leading quark bilinears 
that have non-vanishing overlap with $\langle V_2|$ are 
\begin{equation}
\bar\chi \!\not\!n_- (1\mp \gamma_5)\chi,\quad
\bar\chi \!\not\!n_- \gamma_\perp^\mu (1\pm \gamma_5)\chi.
\end{equation}
The subscript $\perp$ denotes projection of a Lorentz vector on the
plane transverse to the two light-cone vectors $n_\mp$.  The first
operator overlaps only with the longitudinal polarisation state of
$V_2$, the second only with a transverse vector meson.  However, the
second operator is not generated by the $V-A$ interactions of the
Standard Model, at least up to the one-loop level.  Hence the
transverse amplitudes arise from power-suppressed operators
$\mathcal{O}_\perp = [\bar\chi D_\perp \chi] [\bar\xi h_v]$. (In terms
of the light-cone projector (\ref{eq:Mperp}) this statement implies
that the leading term in the first line does not contribute for $V-A$
interactions, leaving the twist-3 terms in the second and third line.
Since $\mathcal{O}_\perp$ contains transverse momentum derivatives,
one must keep the transverse momenta of partons collinear to $V_2$;
this explains why the transverse-momentum derivative terms in the
projector are required.) The left-handedness of the weak interactions
implies that operators of this form contribute only to the
negative-helicity amplitude. The positive-helicity amplitude appears
first in yet higher-dimensional operators such as 
$[\bar\chi D_\perp \chi] [\bar\xi D_\perp h_v]$. To match to such
operators one must keep the transverse momentum of the quark lines
collinear to both mesons non-zero. Such a calculation has not yet been
done, and therefore all calculations of the positive-helicity
amplitude in the literature must be regarded as incomplete. It is even
possible that for the positive-helicity amplitude no useful
factorisation formula holds even for the non-spectator-scattering
terms in (\ref{eq:QCDFformula}).

It follows that the positive-helicity amplitude is power-suppressed
relative to the negative one, and should be set to zero in the absence
of any consistent calculation of this power correction.  Within this
approximation, $\bar{\Amp}_+=\Amp_-=0$, there are only two rather than
four independent polarisation observables, since
\begin{equation}
\label{pluszero}
  f_\parallel = f_\perp, \qquad\qquad
  \phi_\parallel = \phi_\perp.
\end{equation}
Similarly identities for the corresponding CP asymmetries hold. It
should be noted that these identities are non-trivial consequences of
the $V-A$ nature of the weak interactions \emph{and} of factorisation,
and it is therefore worthwhile to test them experimentally.

In our analysis we proceed as follows: we assume the
naive-factorisation expression for the positive-helicity amplitude
$\bar{\Amp}_+$ and allow the form factor to vary within the range
$F_+^{\B\to V_1} = 0\pm 0.06$. Thus, we allow a small variation of
$\bar{\Amp}_+$ around zero to estimate the error from neglecting this
power correction. We note that QCD sum rule results for the form
factors give $F_+$ values consistent with zero
\cite{Ball:1998kk,Ball:2004rg}.

\subsubsection{Violation of the amplitude hierarchy}
\label{subsec:violation}

In the previous paragraphs we explained the origin of the amplitude 
hierarchy (\ref{eq:hierarchy}). However, when electromagnetic effects are 
included, a transverse polarisation amplitude can be 
generated by a short-distance transition to a vector meson and 
a photon with small virtuality which subsequently converts to a 
vector meson \cite{BRYEWP}. This transition is enhanced by a 
factor $(m_B/\LQCD)^2$ due to the large photon propagator, resulting 
in the parametric relation 
\begin{equation}
  \label{eq:emhierarchy}
  \bar{\Amp}_0 : \bar{\Amp}_- : \bar{\Amp}_+ = 
  1 : \frac{\alphaEM m_b}{\LQCD} : \alphaEM.
\end{equation}
Thus, formally, the negative-helicity amplitude is leading in 
the heavy-quark limit. Technically, this is related to the 
existence of the operator  $[A_{\gamma_\perp}] [\bar\xi h_v]$, 
which contributes only to the transverse amplitude, and 
which is enhanced relative to $[\bar\chi\chi] 
[\bar\xi h_v]$. In the Standard Model the leading effect of 
this type involves the electromagnetic-dipole operator 
in the effective Hamiltonian. As a consequence 
the colour-allowed electroweak penguin amplitude $\alpha_{3\, \rm\EW}^{p-}$ 
is completely different from its naive-factorisation value. 
Our calculations include this contribution, which has already 
been discussed specifically in \cite{BRYEWP}.

\subsubsection{Penguin weak annihilation}
\label{sec:penweakannh}

Weak annihilation is a power correction not included in
(\ref{eq:QCDFformula}), since it does not factorise due to endpoint
divergences in the convolution integrals.  The effect is often
estimated by a parameterisation suggested in \cite{BBNS2001}, where
the endpoint-divergences are regulated by a cut-off. In this model one
finds that the most important annihilation effect is a penguin
annihilation amplitude that is phenomenologically indistinguishable
from the QCD penguin amplitude. These general observations also hold
for the transverse polarisation amplitudes. In particular, the weak
annihilation contribution to the negative-helicity amplitude is a
power correction relative to the leading, factorisable contributions
to this amplitude.  Yet, as found in \cite{Kagan:2004uw}, the effect
is numerically much larger than in $\B\to PP, PV$ decays, and perhaps
so large that a theoretical calculation of the negative-helicity QCD
penguin amplitude is no longer possible.

To explain this point, we consider the 
QCD penguin amplitude
\begin{equation}
P^h = A^h_{V_1 V_2} \left[\alpha_4^h+\beta_3^h\right],
\end{equation}
and compare the $h=0$ and $h=-$ amplitudes. Here $\alpha_4^h$ are the
QCD penguin contributions, and $\beta_3^h$ the penguin annihilation
contributions. For the longitudinal amplitude $\beta_3^0$ is
$\LQCD/m_b$ suppressed relative to $\alpha_4^0$, but it turns out that
numerically the largest effect arises from a $(\LQCD/m_b)^2$ term,
which has a large colour factor and Wilson coefficient. This
particular contribution is not suppressed by the extra factor of
$\LQCD/m_b$ in the negative-helicity amplitude.  One finds $ A^-_{V_1
  V_2} \beta_3^- \approx A^0_{V_1 V_2} \beta_3^0$, while $ A^-_{V_1
  V_2} \alpha_4^- \ll A^0_{V_1 V_2} \alpha_4^0$ due to the power
suppression of $ A^-_{V_1 V_2}$ with respect to $ A^0_{V_1 V_2}$.
Thus, relative to $\alpha_4^h$ the numerical effect of $\beta_3^h$ is
a factor of $m_b/\LQCD$ larger for the negative-helicity amplitude
than for the longitudinal amplitude (but still power-suppressed since
the suppression was $(\LQCD/m_b)^2$ for $h=0$).

The following numerical estimates illustrate this point. We consider 
the $p=c$, $h=0,-$ helicity amplitudes for $\rho K^*$, and also the $\pi K$ 
amplitude for comparison. The imaginary parts of the amplitudes 
are neglected, since they are not important for this discussion. We then 
find 
\begin{equation}
  \begin{aligned}
    & \alpha_4^c(\pi \bar K)+\beta_3^c(\pi \bar K) 
      &= \;\; -0.09 - \big\{0.02 \, [-0.01,0.05]\big\}, \\[0.2cm]
    & \alpha_4^{c0}(\rho \bar K^*)+\beta_3^{c0}(\rho \bar K^*) 
      &= \;\; -0.03 - \big\{0.00 \, [-0.00,0.00]\big\}, \\[0.2cm]
    & \alpha_4^{c-}(\rho \bar K^*)+\beta_3^{c-}(\rho \bar K^*)
      &= \;\; -0.05 - \big\{0.03 \, [-0.04,0.10]\big\}.
  \end{aligned}
\end{equation}
The numbers in curly brackets refer to $\beta_3^{p,h}$. The first
number in brackets is the default value, while the interval provides
the range allowed by the parameterisation adopted in \cite{BBNS2001}.
We observe a (presumably accidental) cancellation in the longitudinal
annihilation amplitude. What is significant is the difference in the
range of the interval relative to $\alpha_4^{c,h}$ for the
negative-helicty $VV$ amplitude vs.~the $PP$ amplitude. In particular,
the annihilation contribution $\beta_3^{c-}$ may be significantly
larger than the QCD penguin amplitude $\alpha_4^{c-}$ for the
negative-helicity amplitude. We can also compare the $h=-$ and $h=0$
amplitudes,
\begin{equation}
  \frac{P^-}{P^0}
    \approx \frac{A^-_{\rho K^*}}{A^0_{\rho K^*}} 
            \frac{\alpha_4^{c-}+\beta_3^{c-}}{\alpha_4^{c,0}}
    \approx \frac{0.05+[-0.04,0.10]}{0.12},
\end{equation}
where we used $A^-_{\rho K^*}/A^0_{\rho K^*}\approx 1/4$.  This shows
that $P^-$ could be as large as $P^0$, if annihilation is maximal.
Thus, for penguin-dominated decays, a longitudinal polarisation
fraction $f_L$ around 0.5 is not ruled out.

Let us summarise these and a few further observations on the role of 
weak annihilation in $\B\to VV$ decays:
\begin{itemize}
\item[1)] The annihilation contribution to the longitudinal penguin
  amplitude is small, perhaps due to an accidental cancellation.
\item[2)] The annihilation contribution to the negative-helicity
  penguin amplitude can (but need not) be very large, possibly leading
  to significant transverse polarisation in penguin-dominated decays.
\item[3)] No such enhancement is observed for the annihilation
  contribution to the tree amplitudes, hence tree-dominated decays
  should be predominantly longitudinally polarised.
\item[4)] We also calculated the weak annihilation contribution to the
  positive-helicity amplitude, and find that the large contribution to
  $\beta_3^+$ is absent. Hence there is no evidence for large
  corrections to (\ref{pluszero}) even for penguin-dominated decays.
\end{itemize}
It should be clear that these statements assume that the
parameterisation adopted in
\cite{BBNS2001} reproduces correctly the qualitative features 
of the weak annihilation amplitudes.

\section{Input and overview}
\label{sec:inputs}

\subsection{Input parameters}
\label{sec:inppar}

\begin{table}[t]
  \begin{center}
    \begin{tabular}{ccccc}
      \toprule
      \multicolumn{5}{l}{\em QCD scale and running quark masses [GeV]}
      \\
      $\LQCD=\Lambda_{\overline{\mathrm{MS}}}^{(5)}$ & 
      $m_b(m_b)$ &
      $m_c$ & 
      $m_s(2\,\text{GeV})$ &
      $m_q/m_s$ \\
      \midrule
      0.225 & 
      4.2 &
      $1.3\pm 0.2$ &
      $0.09\pm 0.02$ &
      0.0413 \\
      \bottomrule\\[-0.3cm]
      \multicolumn{5}{l}{\em CKM parameters}
      \\
      $\lambda$ &
      $|V_{cb}|$ &
      $|V_{ub}/V_{cb}|$ &
      $\gamma$ &
      \\
      \midrule
      $0.225$ &
      $0.0415\pm 0.0010$ &
      $0.085^{+0.025}_{-0.015}$ &
      $(70\pm 20)^\circ$ &
      \\
      \bottomrule\\[-0.3cm] 
      \multicolumn{2}{l}{\em $\B$ meson parameters}
      &
      $\Bm$ &
      $\Bdq$ &
      $\Bsq$ 
      \\
      \multicolumn{1}{l}{lifetime} &
      $\tau$ [ps] &
      1.64 &
      1.53 &
      1.46      
      \\
      \multicolumn{1}{l}{decay constant} &
      $f_{\B}$ [MeV] &
      \multicolumn{2}{c}{$210\pm 20$} &
      $240\pm 20$
      \\
      &
      $\lambda_B$ [MeV] &
      \multicolumn{2}{c}{$200^{+250}_{-0}$} &
      $200^{+250}_{-0}$ 
      \\
      \bottomrule\\[-0.3cm] 
      \multicolumn{5}{l}{\em Light meson decay constants and
       Gegenbauer moments}
      \\
      &
      $\prho$ &
      $\Kbar^*$ &
      $\pomega$ &
      $\pphi$
      \\
      \midrule
      $f$ [MeV] &
      $209 \pm 1$ &
      $218 \pm 4$ &
      $187 \pm 3$ &
      $221 \pm 3$
      \\
      $f^{\perp}$ [MeV] &
      $150 \pm 25$ &
      $175 \pm 25$ &
      $150 \pm 25$ &
      $175 \pm 25$
      \\
      $\alpha_1$, $\alpha_{1,\perp}$ &
      0 &
      $0.06 \pm 0.06$ &
      0 &
       0
      \\
      $\alpha_2$, $\alpha_{2,\perp}$&
      $0.1 \pm 0.2$ &
      $0.1 \pm 0.2$ &
      $0.0 \pm 0.3$ &
      $0.0 \pm 0.3$
      \\
      \bottomrule
    \end{tabular}
     \vspace{\baselineskip}
    \begin{tabular}{cccccc}
      &&&&&\\[-0.4cm]
      \multicolumn{6}{l}{\em Form factors for vector mesons at $q^2=0$}
      \\
      &
      $\B\to\prho$ &
      $\B\to \K^*$ &
      $\B\to\pomega$ &
      $\B_s\to\Kbar^*$ &
      $\B_s\to\pphi$
      \\
      \midrule
      $A_0$ &
      $0.30^{+0.07}_{-0.03}$ &
      $0.39\pm 0.06$ &
      $0.25^{+0.07}_{-0.03}$ &
      $0.33\pm 0.05$ &
      $0.38^{+0.10}_{-0.02}$
      \\[0.05cm]
      $F_-$ &
      $0.55\pm 0.06$ &
      $0.68\pm 0.07$ &
      $0.50\pm 0.05$ &
      $0.53\pm 0.05$ &
      $0.65^{+0.14}_{-0.00}$
      \\[0.05cm]
      $F_+$ &
      $0.00\pm 0.06$ &
      $0.00\pm 0.06$ &
      $0.00\pm 0.06$ &
      $0.00\pm 0.06$ &
      $0.00\pm 0.06$
      \\
      \bottomrule
    \end{tabular}
 \end{center}
  \caption{Summary of theoretical input parameters. All
    scale-dependent quantities refer to $\mu=2$\,GeV unless indicated
    otherwise. $m_q=(m_u+m_d)/2$.}
  \label{tab:inputs}
\end{table}
The values of the Standard Model and hadronic input parameters are
listed in Table~\ref{tab:inputs}.  When we compare $B\to VV$ modes to
decays with pions in the final state, the additional pion parameters
are $f_\pi=131\,$MeV, $\alpha_2(2\,\mbox{GeV})=0.2\pm 0.15$,
$f_+^{B\pi}(0)=0.25\pm 0.05$.  (The light quark mass values reported
in the Table are only needed for the computation of the pion decay
amplitudes.)  Relative to the analysis of $PV$ final
states~\cite{BN2003} we have implemented several minor parameter
modifications (Wolfenstein parameter $\lambda$, $|V_{cb}|$, $B$ meson
lifetimes and decay constants, Gegenbauer moments of light-meson
light-cone distribution amplitudes), which reflect new measurements or
improved calculations, but individually have little impact on the
calculation of non-leptonic decay amplitudes.  A more important change
concerns the treatment of $|V_{ub}|$ and the $B$ meson parameter
$\lambda_B$, where we (roughly) stick to the same ranges as before,
but choose smaller default values. These values lead to a good
agreement of theoretical calculations with the observed $B\to\pi\pi$
transitions as already noted in~\cite{BN2003}. Regarding the 
value of $|V_{ub}|$ we note that our default value correspnds to 
the one that is favoured by exclusive semi-leptonic $b\to u$
transitions, which is 
smaller than the one from the inclusive decays (see \cite{Ball:2006jz}
for the most recent discussion). We shall see below that the
tree-dominated $B\to\rho\rho$ modes also support this small value
unless the $B\to\rho$ form factors are unacceptably small, 
so all exclusive decays (semi-leptonic, non-leptonic PP, PV and VV 
modes) seem to consistently favour small $|V_{ub}|$. The default 
value we adopt for $\lambda_B$ is significantly smaller than 
the value obtained from QCD sum rule 
calculations~\cite{Braun:2003wx,Khodjamirian:2005ea}, which we use to 
define the upper limit of this parameter's range.  Some of the
longitudinal and negative helicity $B\to V$ form factors have changed
considerably since the publication of~\cite{BN2003} due to the update
of the QCD sum rule calculation~\cite{Ball:2004ye}. Our new values
follow~\cite{Ball:2004ye}, but in some cases a smaller form factor is
adopted to improve the description of data. The smaller values are
compatible with~\cite{Ball:2004ye} within theoretical errors; in these
cases, however, the theoretically allowed parameter range becomes
asymmetric around the default value. The positive-helicity form
factors are set to $0.00\pm 0.06$ (see
Sect.\ref{subsubsec:positiveHelicity}). The renormalization scales are
treated as in~\cite{BN2003} and $\mu$ is varied from $m_b/2$ to $2
m_b$.  The Wilson coefficients $C_i$ are tabulated in \cite{BBNS2001}.

In addition to the well-defined hadronic parameters, the predictions
of QCD factorisation depend on the model parameters $X_H$ and $X_A$,
$X_L$ (see \cite{BBNS2001} and the Appendix for their definition). In
contrast to all previous QCD factorisation calculations there is
model-dependence even at leading power in the heavy quark expansion
for the transverse amplitudes due to an endpoint divergence in
spectator scattering (see Sect.~\ref{sec:spectscat}). This is
parameterised by
\begin{equation}\label{XHparam}
   X_H = \left( 1 + \varrho_H\,e^{i\varphi_H} \right)
   \ln\frac{m_B}{\Lambda_h} \,; \qquad \Lambda_h=0.5\,\mbox{GeV} \,,
\end{equation}
with $\varrho_H=0$ by default, a range defined by $\varrho_H \le 1$,
and an arbitrary phase $\varphi_H$.  $X_A$ related to weak
annihilation is defined in the same way, but here we use
$\varrho_A=0.6\,e^{-i \,40^\circ}$ by default.  The motivation for
this choice will be explained in the context of penguin-dominated
decays. In $B\to VV$ decays the parameter $X_A$ is only relevant for
the negative-helicity penguin amplitudes due to the near-cancellation
of longitudinal weak annihilation (see Sect.~\ref{sec:penweakannh}).
For completeness we note that $X_L$ is evaluated by an equation
similar to (\ref{XHparam}) with $\ln(m_B/\Lambda_h)$ replaced by
$m_B/\Lambda_h$, but $X_L$ is never numerically relevant in the
amplitude calculation.

In the end all parameter (Standard Model, hadronic, model)
uncertainties are added in quadrature, except for the CKM parameters,
which are separated, because the dependence on the CKM parameters
$|V_{ub}|$ and $\gamma$ is interesting: if larger than the hadronic
error for some observable, this observable may be useful to determine
$|V_{ub}|$ or $\gamma$. In general, the first ``error'' on a quantity
will provide the dependence on CKM parameters; the second gives the
``theoretical uncertainty''.

\subsection{Flavour amplitudes}

\begin{table}[t]
  \renewcommand{\arraystretch}{1.1}
  \centering
  \begin{tabular}{cccc}
    \toprule
    Parameter & $h=0$ & $h=-$ \\
    \hline
    &&\\[-0.3cm]
    $\alpha_1(\rho\rho)$ & 
    $\phantom{-}0.94^{+0.10}_{-0.08} + (\phantom{-}0.02^{+0.06}_{-0.06})i$ &
    $\phantom{-}1.14^{+0.30}_{-0.30} + (\phantom{-}0.04^{+0.29}_{-0.29})i$
    \\[0.3cm]
    $\alpha_2(\rho\rho)$ & 
    $\phantom{-}0.31^{+0.20}_{-0.25} + (-0.08^{+0.15}_{-0.15})i$ &
    $-0.19^{+0.77}_{-0.76} + (-0.17^{+0.75}_{-0.75})i$\\[0.3cm]
    $\beta_1(\rho\rho)$ & 
    $\phantom{-}0.05^{+0.03}_{-0.06} + (-0.02^{+0.06}_{-0.03})i$ &
    $\phantom{-}0.01^{+0.01}_{-0.01} + (\phantom{-}0.00^{+0.01}_{-0.01})i$
    \\[0.3cm]
    $\beta_2(\rho\rho)$ & 
    $-0.02^{+0.02}_{-0.01} + (\phantom{-}0.01^{+0.01}_{-0.03})i$ &
    $\phantom{-}0.00^{+0.00}_{-0.00} + (\phantom{-}0.00^{+0.00}_{-0.00})i$
    \\[0.2cm]\hline
    &&\\[-0.3cm]
    $\alpha_3^p(\bar K^*\phi)$ & 
    $\phantom{-}0.003^{+0.004}_{-0.004} + (-0.001^{+0.003}_{-0.003})i$ &
    $-0.005^{+0.014}_{-0.014} + (-0.001^{+0.014}_{-0.014})i$
    \\[0.3cm]
    $\alpha_4^u(\bar K^*\phi)$ & 
    $-0.024^{+0.004}_{-0.005} + (-0.015^{+0.004}_{-0.005})i$ &
    $-0.049^{+0.017}_{-0.018} + (-0.017^{+0.015}_{-0.016})i$
    \\[0.3cm]
    $\alpha_4^c(\bar K^*\phi)$ & 
    $-0.033^{+0.007}_{-0.007} + (-0.011^{+0.005}_{-0.004})i$ &
    $-0.047^{+0.017}_{-0.017} + (-0.002^{+0.015}_{-0.015})i$
    \\[0.3cm]
    $\beta_3^p(\bar K^*\phi)$ & 
    $-0.000^{+0.002}_{-0.002} + (\phantom{-}0.001^{+0.001}_{-0.002})i$ &
    $-0.031^{+0.066}_{-0.065} + (\phantom{-}0.031^{+0.045}_{-0.105})i$
    \\[0.3cm]
    $\beta_4^p(\rho\rho)$ & 
    $-0.005^{+0.006}_{-0.004} + (\phantom{-}0.002^{+0.003}_{-0.006})i$ &
    $-0.001^{+0.001}_{-0.001} + (\phantom{-}0.000^{+0.001}_{-0.001})i$
    \\[0.2cm]\hline
    &&\\[-0.3cm]
    $\alpha_{3,\rm EW}^p(\bar K^*\rho)$ & 
    $-0.008^{+0.001}_{-0.001} + (-0.000^{+0.000}_{-0.000})i$ &
    $\phantom{-}0.015^{+0.004}_{-0.003} + (-0.003^{+0.002}_{-0.002})i$
    \\[0.2cm]
    $\alpha_{4,\rm EW}^p(\bar K^*\phi)$ & 
    $-0.002^{+0.002}_{-0.002} + (\phantom{-}0.001^{+0.001}_{-0.001})i$ &
    $\phantom{-}0.002^{+0.007}_{-0.007} +
    (\phantom{-}0.001^{+0.007}_{-0.007})i$\\[0.1cm]
    \bottomrule
  \end{tabular}
  \caption{Overview of longitudinal and negative-helicity amplitude 
  parameters. \label{tab:ampratios} }
\end{table}

The helicity decay amplitudes such as the example (\ref{ampexample})
are composed of CKM factors, the factorisable coefficients
(\ref{eq:Aexpl}) and the ``flavour amplitudes'' $\alpha_i^h$,
$\beta_i^h$. The flavour amplitudes correspond to the colour-allowed
(colour-suppressed) tree amplitude $\alpha_1^h$ ($\alpha_2^h$), the
QCD penguin amplitudes ($\alpha_4^{p,h}$), the QCD singlet-penguin
amplitudes ($\alpha_3^{p,h}$), and the colour-allowed
(colour-suppressed) electroweak penguin amplitudes $\alpha_{3\rm
  EW}^{p,h}$ ($\alpha_{4,\rm EW}^{p,h}$). There also exist tree
annihilation $\beta_{1,2}^{h}$, penguin annihilation
$\beta_{3,4}^{p,h}$ and electroweak penguin-annihilation amplitudes
$\beta_{3,4,\rm EW}^{p,h}$. The result of our calculation of these
amplitudes (except for the irrelevant electroweak annihilation
amplitudes) is summarised in Table~\ref{tab:ampratios}.  Most of the
later analysis of branching fractions, CP asymmetries and polarisation
observables can be reproduced by inserting these numerical estimates
into the expressions for the decay amplitudes in terms of flavour
parameters in the appendix of~\cite{BN2003}. The flavour parameters
depend on the final and initial state, but this dependence is rather
small and may be ignored for rough estimates. An exception is the
negative-helicity electroweak-penguin amplitude, since the
power-enhanced electromagnetic contribution depends quadratically on
the light-meson mass. The Table gives the numbers for the $\rho K^*$
final states, where the electroweak penguin amplitudes have the most
significant effects \cite{BRYEWP}.

Let us point out the most important features of the $B\to VV$
amplitudes related to the general discussion in the previous section.
The longitudinal QCD penguin amplitude $\alpha_4^{p,0}$ is rather
small, similar to the $VP$ or $PV$ penguin amplitudes.  However, for
$VV$ the QCD penguin annihilation amplitude $\beta_3^{p,0}$ is
strongly suppressed, and irrelevant, in marked difference to the case
of $B\to PV$ decays. A striking result for the negative-helicity
amplitudes is the value and large uncertainty of the colour-suppressed
tree amplitude, and to some extent even of the colour-allowed tree
amplitude.  This reflects the non-factorisation of
spectator-scattering. The same effect is also responsible for a larger
uncertainty and, possibly, large value of the QCD singlet-penguin
amplitude, which may therefore be relevant to the $\phi K^*$ modes. As
has already been discussed in some detail, the QCD penguin
annihilation amplitude $\beta_3^{p,-}$ is large, perhaps larger than
$\alpha_4^{p,-}$, which is evident from the Table. Finally we note
that the negative-helicity electroweak penguin amplitude
$\alpha_{3,\rm EW}^{p,-}$ has a different sign from the corresponding
longitudinal amplitudes, which is a consequence of the additional,
power-enhanced electromagnetic contribution.

\subsection{Classification of decay modes}

We conclude this overview section with a classification of the 
total of 34 $\Bm$, $\Bdq$ and $\Bsq$ decay channels into two light 
vector mesons according to the reliability of the calculation 
of various observables. This classification is motivated by 
the observation that the
peculiarities of the transverse-helicity amplitude calculation
analysed in Section~\ref{subsec:transverseAnatomy} severely limit the
reliability of QCD factorisation for many observables. In effect, 
of all the transverse amplitudes, only the colour-allowed 
tree and electroweak penguin amplitudes can be calculated 
with some accuracy.  
\begin{itemize}
\item \emph{Colour-allowed tree-dominated $\Delta D=1$ decays.}  Most
  observables are amenable to calculation, the exception being
  CP~asymmetries of polarisation observables, which always involve
  transverse QCD penguin amplitudes. Only four decays belong to this
  class, namely $\Bdq\to\rho^+\rho^-$, $\Bm\to\rho^-\rho^0$,
  $\Bm\to\rho^-\omega$, and $\Bsq\to\rho^- K^{*+}$.
\item \emph{Colour-suppressed tree-dominated $\Delta D=1$ decays.} The
  non-factorisation of spectator scattering in the transverse
  amplitude precludes a reliable calculation of polarisation
  observables for these decays. However, if the longitudinal amplitude
  is still dominant, predictions for the CP-averaged branching
  fractions and $A_{\CP}$ can be obtained, and the longitudinal
  polarisation fraction is expected to be close to 1. The five modes
  $\Bdq\to\rho^0\rho^0$, $\Bdq\to\rho^0\omega$, $\Bdq\to\omega\omega$,
  $\Bsq\to\rho^0 K^{*0}$, and $\Bsq\to\omega K^{*0}$ fall into this
  category.
\item \emph{Penguin-dominated decays.} In these modes, no polarisation
  observables can be calculated reliably from theory alone because of
  penguin weak-annihilation effects. As transverse and longitudinal
  contributions cannot be excluded to be of similar magnitude, even
  branching fraction and CP asymmetry predictions will suffer large
  uncertainties. The eleven $\Delta S=1$ modes $B\to \rho \bar K^*,
  \omega \bar K^*, \phi \bar K^*$, $\Bsq\to K^*\bar K^*, \phi\phi$
  with branching fractions in the upper $10^{-6}$ range, and the three
  $\Delta D=1$ modes $\Bdq\to K^{*0}\bar K^{0*}$, $\Bm\to K^{*0}
  K^{*-}$, $\Bsq\to\phi K^{*0}$ with small branching fractions belong
  to this class.
\item \emph{Electroweak or QCD flavour-singlet penguin-dominated
    decays.} These decays are expected to have very small branching
  fractions. They are difficult to predict, if the QCD flavour-singlet
  penguin amplitude plays a role. The five decays $\Bm\to \rho^-\phi$,
  $\Bdq\to\rho^0\phi$, $\Bdq\to\omega\phi$, and $\Bsq\to\rho^0\phi$,
  $\Bsq\to\omega\phi$ belong to this class. The $\Bsq$ decays in this
  class may exhibit a significant, perhaps dominant, contribution from
  the doubly CKM-suppressed, colour-suppressed tree amplitude.
\item \emph{Pure weak annihilation decays.} The six decays falling
  into this category, namely $\Bdq\to K^{*-} K^{*+}$,
  $\Bdq\to\phi\phi$, $\Bsq\to \rho^+\rho^-$, $\Bsq\to \rho^0\rho^0$,
  $\Bsq\to \rho^0\omega$, and $\Bsq\to\omega\omega$, are completely
  annihilation model-dependent, and only rough estimates of their
  branching fractions can be given.
\end{itemize}
As we proceed with our analysis, we will discuss these categories in
order for the tree-dominated and penguin-dominated decays. 

\section{\boldmath Tree-dominated decays}
\label{sec:tree}

The four tree-dominated colour-allowed modes are among the 
few $B\to VV$ decays that can be reliably calculated in QCD 
factorisation.  They fully respect the helicity
amplitude hierarchy (\ref{eq:hierarchy}) and should exhibit 
predominantly longitudinal polarisation.  Power corrections have
limited impact, and the main sources of theoretical uncertainties are 
$V_{ub}$ and form factors. Penguin amplitudes are small, implying 
small direct CP asymmetries and the prospect of a precise 
determination of $\sin 2 \alpha$
from time-dependent $\Bdq\to\prho^+\prho^-$ studies. On the contrary, 
the colour-suppressed tree-dominated decays have much smaller 
branching fractions, and the theoretical calculations are 
limited by large uncertainties in the colour-suppressed 
amplitude $\alpha_2$, in particular in the transverse amplitude, 
where non-factorisation of transverse spectator scattering
can spoil the predictions. In this section we quantify these 
expectations.  

\subsection{Branching fractions and direct CP asymmetries}

We present the CP-averaged branching ratios and 
direct CP asymmetries in Table~\ref{tab:treeBrAvACP}.  

\begin{table}
  \centering
  \let\oldarraystretch=\arraystretch
  \renewcommand*{\arraystretch}{1.2}
  \begin{tabular}{llllll}
    \toprule
        & \multicolumn{2}{c}{$\BrAv$ / $10^{-6}$}
        &
        & \multicolumn{2}{c}{$A_{\CP}$ / percent}
    \\
    \cmidrule(l){2-3}\cmidrule{5-6}
        & Theory
        & Experiment
        &
        & Theory
        & Experiment
    \\
    \midrule
    $\B^-\to\rho^-\rho^0$
        & $18.8^{+0.4}_{-0.4}{}^{+3.2}_{-3.9}$ $(\ast)$
        & $18.2 \pm 3.0$
        &
        & $\phantom{-}0^{+0}_{-0}{}^{+0}_{-0}$
        & $-8 \pm 13$
        \\
    $\Bdq\to\rho^+\rho^-$
        & $23.6^{+1.7}_{-1.9}{}^{+3.9}_{-3.6}$ $(\ast)$
        & $23.1^{+3.2}_{-3.3}$
        &
        & $-1^{+0}_{-0}{}^{+4}_{-8}$
        & $+11 \pm 13$
        \\
    $\Bdq\to\rho^0\rho^0$
        & $\phantom{0}0.9^{+0.6}_{-0.3}{}^{+1.9}_{-0.9}$
        & $1.07\pm 0.38$
        &
        & $+28^{+5}_{-7}{}^{+53}_{-29}$
        & n/a
        \\\addlinespace
    $\B^-\to\omega\rho^-$
        & $12.8^{+1.1}_{-1.3}{}^{+2.0}_{-2.4}$ $(\ast)$
        & $10.6^{+2.6}_{-2.3}$ 
        &
        & $-8^{+3}_{-2} {}^{+5}_{-8}$
        & $+4 \pm 18$
        \\
    $\Bdq\to\omega\rho^0$
        & $\phantom{0}0.2^{+0.1}_{-0.1}{}^{+0.3}_{-0.1}$
        & $<1.5$
        &
        & no prediction
        & n/a
        \\
    $\Bdq\to\omega\omega$
        & $\phantom{0}0.9^{+0.5}_{-0.3}{}^{+1.5}_{-0.9}$
        & $<4.0$
        &
        & $-29^{+9}_{-6}{}^{+25}_{-44}$
        & n/a
        \\\addlinespace
   $\Bsq\to\K^{*+}\rho^-$
        & $25.2^{+1.5}_{-1.7}{}^{+4.7}_{-3.1}$  $(\ast)$
        & n/a
        &
        & $-3^{+1}_{-1}{}^{+2}_{-3}$
        & n/a
        \\
   $\Bsq\to\K^{*0}\rho^0$ 
        & $\phantom{0}1.5^{+1.0}_{-0.5}{}^{+3.1}_{-1.5}$
        & n/a
        &
        & $+27^{+5}_{-7}{}^{+34}_{-27}$
        & n/a
        \\
    $\Bsq\to\K^{*0}\omega$
        & $\phantom{0}1.2^{+0.7}_{-0.3}{}^{+2.3}_{-1.1}$
        & n/a
        &
        & $-34^{+10}_{-7}{}^{+31}_{-43}$
        & n/a
        \\
    \bottomrule
  \end{tabular}
  \let\arraystretch=\oldarraystretch
  \caption{CP-averaged branching fractions and direct CP asymmetries
    of tree-dominated $B\to VV$ decays. Experimental values are taken
    from~\cite{Aubert:2006vt,Aubert:2006sb,Aubert:2006af,Aubert:2006wx,Zhang:2003up,Somov:2006sg}.
    For numbers marked with an asterisk, the dependence on
    $\abs{V_{ub}}$ and the form factor is obtained as described in the
    text. Errors are calculated as described in Section~\ref{sec:inppar}.}
  \label{tab:treeBrAvACP}
\end{table}

It is interesting to note that the experimental data on the $\rho\rho$
branching fractions exhibit a pattern similar to the corresponding
$\pi\pi$ modes. The $\rho^+\rho^-$ and $\rho^-\rho^0$ modes have
nearly equal branching fractions, and the $\rho^0\rho^0$ has a larger
branching fraction than naively expected. For both, pions and $\rho$
mesons, this is attributed to a larger colour-suppressed tree
amplitude. In the factorisation framework this is realized if
spectator scattering is the dominant dynamical mechanism behind the
colour-suppressed tree amplitude. This favours the parameter choice
adopted in Section~\ref{sec:inputs} with small $\lambda_B$ and small
$V_{ub}$ and/or form factors. It is seen from the Table that the
existing data is consistent with the theoretical calculation.

The colour-suppressed decays are easily distinguished in the Table by
their small branching fractions and large relative uncertainties.
These uncertainties are dominated by the parameters for power
corrections (mainly $X_H$ relevant to spectator scattering) and there
is little room for improvement from theory alone. The CKM error on the
branching fractions is dominated by $|V_{ub}|$. Except for
$\Bdq\to\omega\rho^0$, the dependence on $|V_{ub}|$ can be extracted
by assuming that the branching fraction is proportional to
$|V_{ub}|^2$.  It is worth analysing the uncertainties of the
colour-allowed decays in more detail, since they are dominated by
$V_{ub}$ and form factors. Both can be expected to be known more
accurately in the nearer future. Only the longitudinal form factors
$A_0^{B\to M_1}(0)$ are relevant here, since the transverse amplitudes
contribute only a small amount to the branching fraction.  To make
this dependence explicit, we write
\begin{equation}
 \BrAv(\B\to\prho^-\prho^0) = 
      \abs{\frac{V_{ub}}{3.53\cdot 10^{-3}}}^2 \times
      \abs{\frac{A_0^{\B\to\prho}(0)}{0.30}}^2 \times
      \left(18.8^{+0.4}_{-0.4}{}^{+3.2}_{-3.9}\right)\cdot 10^{-6}
\label{ffextract}
\end{equation}
for the $\rho^-\rho^0$, and similarly for the $\rho^+\rho^-$ final
state. For $\B^-\to\omega\rho^-$ and $\Bsq\to\K^{*+}\rho^-$ we extract
the default values of $A_0^{B\to \omega}(0)$ and $A_0^{B_s\to
  K^*}(0)$, respectively. We then quote the number
$18.8^{+0.4}_{-0.4}{}^{+3.2}_{-3.9}$ in Table~\ref{tab:treeBrAvACP},
where the error includes all parameters but only the residual
dependence on $|V_{ub}|$ and the form factor. The bulk dependence can
then be obtained by inserting into (\ref{ffextract}) whatever values of
$|V_{ub}|$ and the form factor one prefers.  We may conclude from the
branching fractions of the colour-allowed decays that a significantly
larger value of $|V_{ub}|$ (than our default value
$|V_{ub}/V_{cb}|=0.085$) is only compatible with data, if all form
factors are substantially below the current QCD sum rule results.

Certain ratios of branching fractions can shed more light on 
the underlying hadronic dynamics. The hadronic uncertainties 
on the two ratios 
\begin{align}
  \frac{\BrAv(\B\to\prho^-\prho^0)}{\BrAv(\B\to\prho^+\prho^-)} 
    &= 0.80^{+0.05}_{-0.04}{}^{+0.25}_{-0.26} 
    &  \text{(exp: $0.79\pm 0.17$)},\hspace*{0.3cm}\\ 
  \frac{\BrAv(\B\to\prho^0\prho^0)}{\BrAv(\B\to\prho^+\prho^-)} 
    &= 0.038^{+0.009}_{-0.007}{}^{+0.090}_{-0.041} 
    &  \text{(exp: $0.046 \pm 0.016$})
\end{align}
are determined almost entirely by spectator scattering such that a 
larger ratio implies that this mechanism is more important. 
On the other hand, 
\begin{equation}
  \frac{\BrAv(\B\to\prho^-\omega)}{\BrAv(\B\to\prho^-\prho^0)} 
   =  \abs{\frac{1.2\,A_0^{\B\to\omega}(0)}{A_0^{\B\to\prho}(0)}}^2 \times
      \left(0.68^{+0.05}_{-0.06}{}^{+0.11}_{-0.09}\right) 
  \qquad 
  \text{(exp: $0.58^{+0.17}_{-0.16}$)}
\end{equation}
provides insight on the 
ratio of the $B\to\omega$ to $B\to\rho$ form factor, as indicated 
by the above dependence on this ratio. More interesting information 
could be obtained from the $\rho\rho$ final states, once the 
semi-leptonic $B\to\rho\ell\nu$ spectrum is measured more 
accurately near $q^2=0$. 

The predicted direct CP asymmetries are either very 
uncertain (colour-suppressed modes), often preferring only 
one or the other sign of the asymmetry, or rather small 
(colour-allowed decays). The small asymmetries for the 
colour-allowed decays follow from the dominance of the 
longitudinal polarisation amplitude combined with the 
smallness of the penguin amplitude. The available measurements are
consistent with small or vanishing asymmetries, but do not 
allow to draw further conclusions at this moment.  

\subsection{\boldmath Longitudinal amplitudes and the determination of 
$\alpha$ ($\gamma$) from $S_L$}

For phenomenological studies it is often convenient to parameterise 
the decay amplitudes by hadronic amplitudes that can be directly 
determined from data. In the limit of isospin symmetry and neglecting 
electroweak penguin contributions, the $\B\to\prho\prho$ amplitude 
system is conventionally written in terms of complex graphical ``tree'',
``colour-suppressed tree'' and ``penguin'' amplitudes, 
\begin{equation}
  \begin{split}
    \sqrt{2}\,\Amp^h_{\Bm\to\prho^-\prho^0}  &= (T^h+C^h)\,\e^{-\I\gamma}, \\
              \Amp^h_{\Bdq\to\prho^+\prho^-} &=  T^h\e^{-\I\gamma} + P^h, \\
            - \Amp^h_{\Bdq\to\prho^0\prho^0} &= C^h\e^{-\I\gamma} - P^h.
  \end{split}
\end{equation}
A similar set of equations applies to the $\B\to\ppi\ppi$ system. 
This amounts to five real hadronic parameters per helicity 
amplitude. Given $\gamma$ they can be extracted from the three 
helicity-specific branching fractions, the direct 
CP asymmetry in $\Bdq\to (\rho^+\rho^-)_h$, and the 
time-dependent CP asymmetry $S_{h}^{\rho\rho}$, and compared to 
theoretical calculations.

We calculate these quantities in QCD factorisation, where the main
contributions to $T$, $C$ and $P$ come from the coefficients
$\alpha_1$, $\alpha_2$ and $\alpha_4^c+\beta_3^c$, respectively. In
the following discussion we will only consider the longitudinal
amplitudes, drop the helicity index and write $C = \abs{T} \times
r_C\,\e^{\I\delta_C}$, $P = \abs{T} \times r_P\,\e^{\I\delta_P}$.  The
results are given in Table~\ref{tab:rhorhopipi}, which also compares
the $\rho\rho$ to the $\pi\pi$ system. The errors in this Table are
from hadronic parameters. $r_C$, $\delta_C$, $\delta_P$ do not depend
on CKM parameters. The uncertainty from $|V_{ub}|$ and $V_{cb}$ can be
included noting that $T$ is proportional to $|V_{ub}|$, $r_P$ to
$|V_{cb}|/|V_{ub}|$. Similar results have been presented in
\cite{BJ2}. Numerical differences arise from a different choice of
input parameters (for instance, here we use the same value
$\varrho_A\,e^{i\varphi_A}=0.6\,e^{-i \,40^\circ}$ for pions and
$\rho$ mesons) and the inclusion of spectator-scattering effects at
next-to-next-to-leading order in \cite{BJ2}. The values reported here
and in \cite{BJ2} provide a good description of all available $\pi\pi$
and $\rho\rho$ observables within uncertainties of the calculation
with the exception of the direct CP asymmetry in $\Bdq\to\pi^+\pi^-$,
which is predicted to be smaller than what is observed (by the BELLE 
experiment).

\begin{table}
  \centering
  \let\oldarraystretch=\arraystretch
  \renewcommand*{\arraystretch}{1.2}
  \begin{tabular}{cll}
    \toprule
       & $\B\to\prho\prho$
       & $\B\to\ppi\ppi$ \\
    \midrule
    $\abs{T}/(10^{-8}\,\mathrm{GeV}^{-1})$ 
       & $4.74^{+1.27}_{-0.68}$
       & $2.15^{+0.58}_{-0.55}$ \\
    $r_C$
       & $0.30^{+0.24}_{-0.26}$
       & $0.57^{+0.34}_{-0.40}$ \\
    $\delta_C$
       & $(-10^{+33}_{-38})^\circ$
       & $(-4^{+22}_{-23})^\circ$ \\
    $r_P$
       & $0.12^{+0.02}_{-0.03}$
       & $0.42^{+0.17}_{-0.15}$ \\
    $\delta_P$
       & $(8^{+14}_{-7})^\circ$
       & $(0^{+26}_{-12})^\circ$ \\
    \bottomrule
  \end{tabular}
  \let\arraystretch=\oldarraystretch
  \caption{Amplitude parameters of the $\rho\rho$ and $\pi\pi$
    systems. Errors do not include CKM parameter uncertainties.}
  \label{tab:rhorhopipi}
\end{table}

It is interesting to understand the difference between 
the $\rho\rho$ and $\pi\pi$ system. The value of $|T|$, which 
controls the absolute magnitude of the colour-allowed 
branching fractions, is larger for $\rho$ mesons, because 
the product $f_\rho A_0^{B\to\rho}(0)$ of decay constant and 
form factor is larger than for pions, see 
Section~\ref{sec:inputs}. The second important difference is 
the smaller penguin-to-tree ratio $r_P$ for $\rho$ mesons, 
which follows from the absence of the power suppressed 
but ``chirally-enhanced'' scalar penguin amplitude. This  
also causes $r_C$ to differ, because $T=\alpha_1+\alpha_4^u+\ldots$, 
and $C=\alpha_2-\alpha_4^u+\ldots$, and explains why 
the branching fraction of $\Bdq\to\prho^+\prho^-$ is about four times 
larger than  $\Bdq\to\pi^+\pi^-$, while those of
$\Bdq\to\prho^0\prho^0$ and  $\Bdq\to\pi^0\pi^0$ are about equal.

We now turn to the determination of $\gamma$ (or $\alpha$) from 
time-dependent CP violation. The two CP asymmetries are defined 
through 
\begin{equation}
  \frac{ \Gamma_L(\Bbar^0(t)\to\prho^+\prho^-)
        -\Gamma_L(\B^0(t)\to\prho^+\prho^-)}%
       { \Gamma_L(\Bbar^0(t)\to\prho^+\prho^-)
        +\Gamma_L(\B^0(t)\to\prho^+\prho^-)}
  = - C_L^{\prho\prho} \cos(\Delta m t)
    + S_L^{\prho\prho} \sin(\Delta m t),
\end{equation}
where $\Delta m>0$ is the mass difference of the two neutral 
$B$ meson mass eigenstates. We obtain 
\begin{equation}
  C_L^{\prho\prho} =  0.027^{+0.007}_{-0.009}{}^{+0.047}_{-0.024} 
\qquad \mbox{(exp: } -0.11\pm 0.13),
\end{equation}
in good agreement with experiment. As emphasized in
\cite{BN2003,BBNS2001}, the asymmetries $S_f$ are particularly suited
to determine the CKM phase in the framework of QCD factorisation,
because hadronic uncertainty enters only in the penguin-correction
term to $S_f$, and the dependence on the strong phase $\delta_P$ comes
through $\cos \delta_P$ in very good approximation. Thus, like in no
other observable, the hadronic uncertainty is much smaller than the
dependence on the CKM phase $\gamma$. This is especially true for the
$\rho\rho$ system, where $r_P$ is small, so that
\begin{equation}
S_L^{\rho\rho} = \sin2\alpha + 2 \,r_P\cos\delta_P
\sin\gamma \cos2\alpha + O(r_P^2)
\end{equation}
with $\alpha=\pi-\beta-\gamma$. 

\begin{figure}
  \centering
  \includegraphics{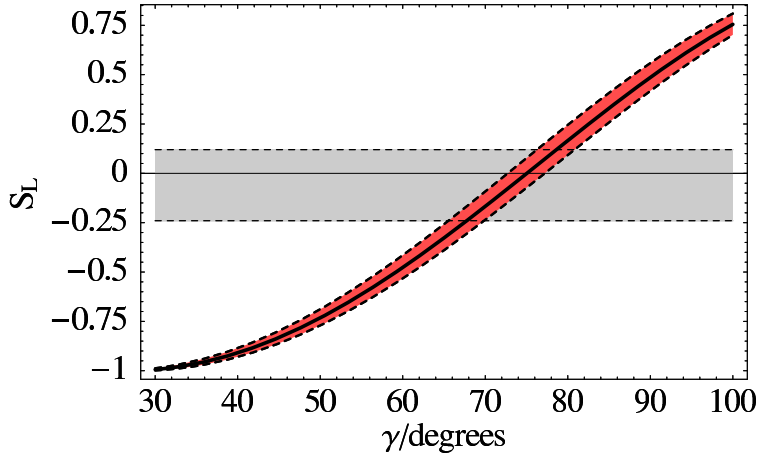}
  \caption{Theoretical result for the longitudinal time-dependent
    CP-asymmetry parameter $S_L^{\prho\prho}$ as a function of the CKM
    angle $\gamma$. The horizontal band indicates the current
    experimental constraint \cite{Aubert:2006af,Abe:2007ez}.
  \label{fig:rhorho_gamma-SL}}
\end{figure}

Figure~\ref{fig:rhorho_gamma-SL} shows 
$S_L^{\prho\prho}$ as a function of $\gamma$ in the range of 
interest. The band quantifies the theoretical uncertainty. From 
the intersection of this band with the measured value 
$-0.06\pm 0.18$~\cite{Aubert:2006af,Abe:2007ez}, and given 
$\beta=(21.2\pm 1.0)^\circ$ \cite{hfag2006,Aubert:2006aq,Chen:2006nk}, 
we obtain 
\begin{equation}
  \gamma = (73.2^{+7.6}_{-7.7})^\circ
\qquad \mbox{or} \qquad 
  \alpha = (85.6^{+7.4}_{-7.3})^\circ,
\end{equation}
where the theoretical error alone is only $\pm 3^\circ$. The value 
of $\gamma$ obtained in this way is remarkably consistent with 
the one from QCD factorisation calculations of 
the $S$-parameters of the $\pi^+\pi^-$ and 
$\pi^\pm\rho^\mp$ final states \cite{BN2003,MBoxford}, 
and presently provides the 
most accurate direct determination of $\gamma$. It is also consistent 
with \cite{BGRS}, where instead of a theoretical calculation of 
$P/T$ one uses SU(3) symmetry to relate the penguin amplitude 
to the longitudinal branching fraction of 
$\Bm\to\rho^- \bar K^{*0}$, and with other determinations of 
$\gamma$ ($\alpha$) from $B\to \rho\rho$ 
decays~\cite{Aubert:2006af,Abe:2007ez}.

\subsection{Polarisation observables}

We now study the transverse-helicity contributions, which manifest 
themselves in polarisation observables. As explained before,
model-dependent effects such as non-factorisation of
spectator-scattering and penguin annihilation can make a strong impact
on transverse amplitudes, and therefore our results suffer from 
larger uncertainties. Here, this specifically concerns
the five colour-suppressed modes.
On the other hand, as polarisation observables like $f_L$ involve
ratios, other uncertainties are often reduced.  Specifically, CKM
factors often drop out approximately, and form factors only enter in
form of ratios (for example transverse/longitudinal), when only one
form factor contribution is present or strongly dominant in an
amplitude.

\begin{table}
  \centering
  \let\oldarraystretch=\arraystretch
  \renewcommand*{\arraystretch}{1.2}
  \begin{tabular}{lllll}
    \toprule
        & \multicolumn{2}{c}{$f_L$ / percent}
        &
        & \multicolumn{1}{c}{$A_{\CP}^0$ / percent}
    \\
    \cmidrule(l){2-3}
    \cmidrule{5-5}
        & Theory
        & Experiment
        &
        & Theory
    \\
    \midrule
    $\B^-\to\prho^-\prho^0$
       & $95.9^{+0.2}_{-0.3}{}^{+3.4}_{-5.9}$
       & $91.2^{+4.4}_{-4.5}$ 
       &
       & $-0^{+0}_{-0}{}^{+0}_{-0}$
       \\
    $\Bdq\to\prho^+\prho^-$
       & $91.3^{+0.4}_{-0.3}{}^{+5.6}_{-6.4}$ 
       & $96.8 \pm 2.3$
       &
       & $-2^{+0}_{-0}{}^{+4}_{-2}$ 
       \\
    $\Bdq\to\prho^0\prho^0$
       & $90^{+3}_{-4}{}^{+8}_{-56}$ 
       & $87\pm 14$
       &
       & $-8^{+2}_{-1}{}^{+59}_{-28}$ 
       \\\addlinespace
    $\B^-\to\pomega\prho^-$
       & $93.7^{+1.1}_{-1.0}{}^{+4.7}_{-8.1}$ 
       & $82 \pm 11$
       &
       & $-2^{+1}_{-0}{}^{+7}_{-6}$ 
       \\
    $\Bdq\to\pomega\prho^0$
       & $49^{+11}_{-11}{}^{+47}_{-23}$ 
       & n/a
       &
       & $+35^{+25}_{-15}{}^{+47}_{-84}$ 
       \\
    $\Bdq\to\pomega\pomega$
       & $93^{+2}_{-4}{}^{+5}_{-22}$ 
       & n/a
       &
       & $+6^{+1}_{-1}{}^{+14}_{-24}$
       \\\addlinespace
    $\Bsq\to\K^{*+}\rho^-$
        & $92.2^{+0.6}_{-0.5}{}^{+5.2}_{-7.5}$ 
        & n/a
        &
        & $-2^{+1}_{-0}{}^{+6}_{-3}$ 
        \\
    $\Bsq\to\K^{*0}\rho^0$
        & $93^{+2}_{-3}{}^{+5}_{-54}$ 
        & n/a
        &
        & $-5^{+1}_{-0}{}^{+49}_{-18}$ 
        \\
    $\Bsq\to\K^{*0}\omega$
        & $93^{+2}_{-4}{}^{+5}_{-49}$ 
        & n/a
        &
        & $+6^{+1}_{-1}{}^{+19}_{-60}$
        \\
    \bottomrule
  \end{tabular}
  \let\arraystretch=\oldarraystretch
  \caption{Longitudinal polarisation fraction and the corresponding CP
    asymmetry for tree-dominated decays. Experimental values are taken
    from~\cite{Aubert:2006vt,Aubert:2006sb,Aubert:2006af,Aubert:2006wx,Zhang:2003up,Somov:2006sg}.}
  \label{tab:fLtree}
  \end{table}

Our results for the longitudinal polarisation fraction and the 
corresponding CP asymmetry are shown in Table~\ref{tab:fLtree}. 
As expected the colour-allowed tree-dominated decay modes are
predicted to have $f_L$ near 1 with errors in the $(5-10)\%$ 
range. Their longitudinal CP asymmetries are predicted not to 
exceed $10\%$. Again the situation is very different for the 
colour-suppressed decays. With the exception of
$\Bdq\to\pomega\prho^0$ there is still a preference for 
significant longitudinal polarisation, but much smaller
values can be obtained within theoretical errors. The large 
downward uncertainty is entirely due to the model-dependence 
in the spectator-scattering contribution to the negative helicity 
amplitude $\alpha_2^-$ (parameterised by $X_H$ in the QCD factorisation 
formalism). The longitudinal CP asymmetries are also very 
uncertain. 

\begin{figure}
  \centering
  \includegraphics[width=11cm]{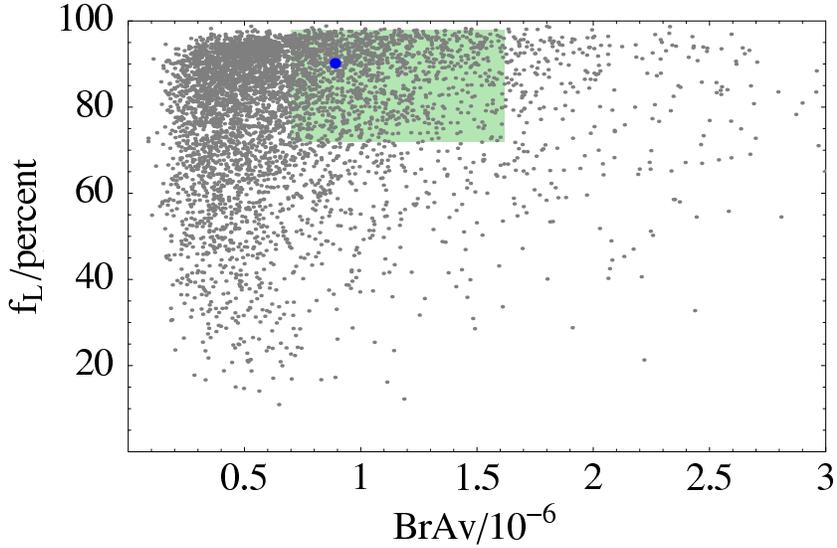}
  \caption{Correlation of branching fraction and longitudinal
    polarisation fraction in $\Bdq\to\prho^0\prho^0$, illustrated
    using 5000 randomly chosen points in our parameter space. The
    shaded area corresponds to the BABAR measurement
    \cite{Aubert:2006wx}, the highlighted point to our default
    values.}
  \label{fig:rho0rho0_BrAv-fL}
\end{figure}

The theoretical predictions of $f_L$ are compatible with the 
present experimental data where available. One notices, however, 
that the pattern of $1-f_L$ for the two colour-allowed 
$\rho\rho$ final states seen by experiment appears to be opposite 
to the theoretical one though perhaps not significantly. We 
therefore calculate
\begin{equation}
  r_{+0} = \frac{1-f_L(\rho^-\rho^0)}{1-f_L(\rho^+\rho^-)} 
         =  0.46^{+0.06}_{-0.04}{}^{+1.63}_{-0.42}
  \qquad \mbox{(exp: } 2.75^{+2.43}_{-2.41}).
\end{equation}
The theoretical upper limit of $r_{+0}\approx
2$ is attained when spectator-scattering is minimal ($X_H=-1$), in
which case $f_L(\rho^-\rho^0)\approx 0.90$ and 
$f_L(\rho^+\rho^-)\approx 0.95$. The branching fraction and 
longitudinal polarisation fraction of the colour-suppressed 
decay $\Bdq\to\rho^0\rho^0$ are theoretically allowed to lie 
within large ranges. To investigate the question whether there 
exist (theoretical) correlations between the two observables, 
we perform a random scan through the theory parameter space. 
The result is displayed in Figure~\ref{fig:rho0rho0_BrAv-fL}. It 
shows that while there is a preference for smaller branching fractions 
than in our default prediction, there is no obvious correlation 
between the branching fraction and $f_L$. 

The remaining six polarisation observables can be taken to be 
$f_\parallel-f_\perp$, $A_{\rm CP}^\parallel-A_{\rm CP}^\perp$, 
and the phase differences between the transverse helicity amplitudes and 
the longitudinal amplitude. As explained in 
Section~\ref{sec:factorisation}, $f_\perp$, 
$A^\perp_{\rm CP}$, $\phi_\perp$, $\Delta\phi_\perp$ 
are expected to be approximately equal to  
$f_\parallel$, $A^\parallel_{\rm CP}$, 
$\phi_\parallel$, $\Delta\phi_\parallel$. We find indeed that 
$f_\parallel-f_\perp$ is always below $2\%$. The last column 
of Table~\ref{tab:angleobservables} quantifies the 
expectation of equal $\phi_\parallel$ and $\phi_\perp$: the 
difference of the two does 
not exceed a few degrees. We should point out, however, that 
this calculation relies on the assumption that the positive-helicity 
is not substantially different from its magnitude in naive 
factorisation. In view of this the smallness of  
$\phi_\parallel-\phi_\perp$ should be interpreted as the statement 
that no concrete dynamical mechanism is known that could 
produce a larger difference. The phase observables 
$\phi_\parallel$ and the corresponding CP asymmetry
$\Delta\phi_\parallel$ are shown in the second and third column of 
Table~\ref{tab:angleobservables}. For most of the colour-suppressed 
decays the theoretical uncertainty is above $270^\circ$, in which 
case we conclude that no useful theoretical prediction is 
possible. For the colour-allowed modes, we obtain reasonably 
accurate results, which could be compared to experiment, once a 
complete angular analysis is performed.

\begin{table}
  \centering
  \let\oldarraystretch=\arraystretch
  \renewcommand*{\arraystretch}{1.2}
  \begin{tabular}{lllc}
    \toprule
       & $\phi_\parallel$ / degrees
       & $\Delta\phi_\parallel$ / degrees
       & $(\phi_\parallel-\phi_\perp)$ / degrees \\
    \midrule
    $\B^-\to\prho^-\prho^0$ 
       & $-5^{+0}_{-0}{}^{+31}_{-32}$
       & $-6^{+2}_{-1}{}^{+2}_{-5}$
       & $\pm 2$
       \\
    $\Bdq\to\prho^+\prho^-$
       & $+1^{+2}_{-2}{}^{+17}_{-17}$
       & $+4^{+1}_{-1}{}^{+9}_{-9}$
       & $\pm 0$
       \\
    $\Bdq\to\prho^0\prho^0$
       & no prediction
       & no prediction
       & $\pm 3$
       \\\addlinespace
    $\B^-\to\pomega\prho^-$
       & $-9^{+2}_{-2}{}^{+34}_{-28}$
       & $+18^{+6}_{-6}{}^{+34}_{-31}$
       & $\pm 3$
       \\
    $\Bdq\to\pomega\prho^0$
       & $-29^{+5}_{-4}{}^{+58}_{-58}$
       & $-45^{+13}_{-17}{}^{+61}_{-37}$
       & $\pm 1$
       \\
    $\Bdq\to\pomega\pomega$
       & no prediction
       & no prediction
       & $\pm 3$
       \\\addlinespace
    $\Bsq\to\K^{*+}\rho^-$
       & $-1^{+3}_{-2}{}^{+21}_{-21}$
       & $+7^{+2}_{-2}{}^{+13}_{-14}$
       & $\pm 0$
       \\
    $\Bsq\to\K^{*0}\rho^0$
       & no prediction
       & no prediction
       & $\pm 2$
       \\
    $\Bsq\to\K^{*0}\omega$
       & no prediction
       & no prediction
       & $\pm 3$
       \\
    \bottomrule
  \end{tabular}
  \let\arraystretch=\oldarraystretch
  \caption{Predictions for other polarisation observables}
  \label{tab:angleobservables}
\end{table}

\boldmath
\section{Penguin-dominated decays}
\label{sec:penguin}
\unboldmath

The 14 decay modes that we study in this section are characterised by 
the dominant role of the colour-allowed QCD penguin amplitude 
$\hat{\alpha}_4^p \equiv \alpha_4^p + \beta_3^p$, which includes a 
penguin-annihilation term. Of these the 11 $\Delta S=1$ modes 
have branching fractions up to $10^5$, and some of them have already 
been studied extensively experimentally including polarisation. 

Due to their common dominant amplitude the theoretical errors 
in this class of decays are common to all representatives. As 
explained in Section~\ref{subsec:transverseAnatomy}, the
negative-helicity penguin amplitude $\hat{\alpha}_4^{p-}$ is 
particularly uncertain due to a potentially large penguin weak annihilation 
contribution \cite{Kagan:2004uw}. In addition, non-factorisation of 
spectator scattering also affects the transverse amplitude 
of final states containing $\omega$ or $\phi$ mesons, 
mostly through the flavour-singlet penguin amplitude 
$\alpha_3^{p-}$. An important issue of the subsequent analysis 
will be whether theoretical calculations are compatible 
with the observation of large transverse polarisation, and 
whether uncertainties can be controlled to the point that useful 
predictions can be made.

\subsection{\boldmath 
The $\B\to\pphi\K^{*}$ system and the transverse penguin
  amplitude}
\label{phikstsec}

\begin{figure}
  \centering
  \includegraphics{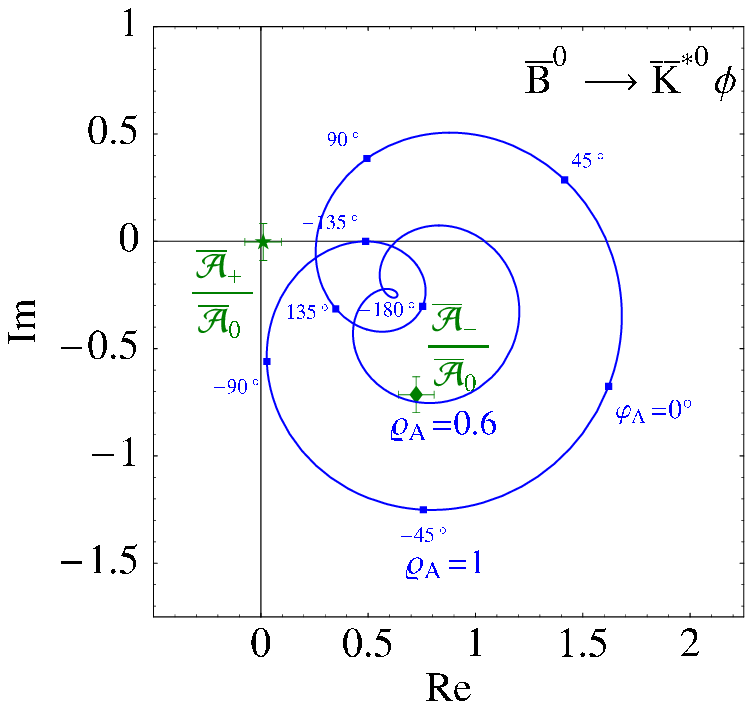}
  \hfill
  \includegraphics{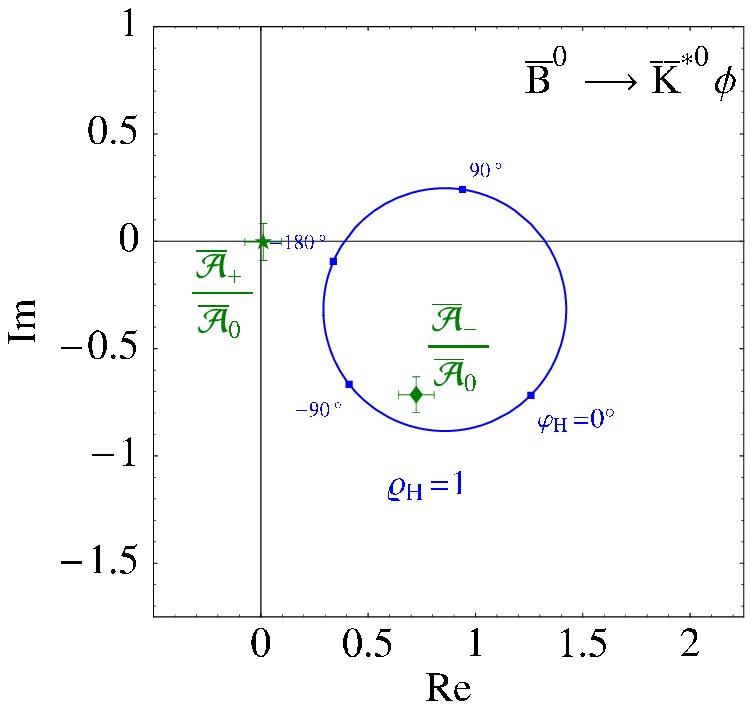}
  \caption{Transverse-to-longitudinal amplitude ratios for
    $\Bdq\to\Kbar^*\phi$. The contours indicate the dependence of the
    negative-helicity theory prediction on annihilation (left) and
    spectator scattering (right) parameters, with all other input
    fixed at central values. Additionally, the values implied by
    current measurements \cite{Aubert:2004xc,newbabar,Chen:2005zv} 
    for both helicities are shown.}
  \label{fig:kstphi_transAmp}
\end{figure}

We begin with a discussion of the $\B\to\pphi\K^{*}$ modes.  A
complete angular analysis is available for $\B\to\pphi\K^{*0}$
\cite{Aubert:2004xc,Chen:2005zv}, which allows us to extract the
complex amplitude ratios $\bar\Amp_\pm/\bar\Amp_0$ from data. This is
shown in Figure~\ref{fig:kstphi_transAmp}, which compares this result
to the theoretical calculation of $\bar\Amp_-/\bar\Amp_0$. (The
experimental result for $\bar\Amp_+/\bar\Amp_0$ is in very good
agreement with the expectation that the plus-helicity amplitude should
be strongly suppressed.) The left plot in the figure shows the
theoretical range from a variation of the uncertainties in weak
annihilation alone (parameter $X_A$), the right plot displays the same
information for spectator scattering (parameter $X_H$).  Since all
values for inside the contour are theoretically allowed for
$\bar{\Amp}_-/\bar{\Amp}_0$, it is evident that theory does not
require the amplitude ratio to be small. While it does not make
accurate predictions, it is natural that penguin-dominated decays
exhibit large transverse polarisation. This is confirmed by comparing
the first column of numbers in Table~\ref{tab:phikst} with the
measurements in the fourth column. We find very good agreement of our
results with data but with very large uncertainties. We also note that
all observables related to the positive-helicity amplitude (the
difference of $\parallel$ and $\perp$ observables) are predicted to be
very small. So are the CP asymmetries, since the doubly CKM-suppressed
amplitude proportional to $\lambda_u^{(s)}$ does not exceed a few
percent. Unless experiments find unexpectedly large values for any of
these observables, the interesting ones are the branching fraction,
$f_L$ and the phase $\phi_\parallel$.

\begin{table}[tp]
  \centering
  \let\oldarraystretch=\arraystretch
  \renewcommand*{\arraystretch}{1.2}
  \begin{tabular}{llllll}
    \toprule
    \multicolumn{2}{l}{Observable} &
    \multicolumn{3}{l}{Theory} &
    Experiment
    \\
    \cmidrule{3-5}
    \multicolumn{2}{l}{}&
    default &
    constrained $X_A$ &
    $\hat{\alpha}_4^{c-}$ from data &
    \\\midrule
    $\BrAv / 10^{-6}$
      & $\pphi\K^{*-}$
        & $10.1^{+0.5}_{-0.5}{}^{+12.2}_{-7.1}$
        & $10.1^{+0.5}_{-0.5}{}^{+7.2}_{-4.8}$
        & $10.4^{+0.5}_{-0.5}{}^{+5.2}_{-3.9}$
        & $9.7\pm 1.5$
        \\
      & $\pphi\Kbar^{*0}$
        & $\phantom{1}9.3^{+0.5}_{-0.5}{}^{+11.4}_{-6.5}$
        & $\phantom{1}9.3^{+0.5}_{-0.5}{}^{+6.7}_{-4.5}$
        & $\phantom{1}9.6^{+0.5}_{-0.5}{}^{+4.7}_{-3.6}$
        & $9.5\pm 0.8$
        \\
        \addlinespace
    $A_{\CP} / \%$
      & $\pphi\K^{*-}$
        & $0^{+0}_{-0}{}^{+2}_{-1}$
        & $0^{+0}_{-0}{}^{+0}_{-0}$
        & $0^{+0}_{-0}{}^{+3}_{-2}$
        & $5\pm 11$
        \\
      & $\pphi\Kbar^{*0}$
        & $1^{+0}_{-0}{}^{+1}_{-0}$
        & $1^{+0}_{-0}{}^{+0}_{-0}$
        & $1^{+0}_{-0}{}^{+2}_{-1}$
        & $-1\pm 6$
        \\
        \addlinespace
    $f_L / \%$
      & $\pphi\K^{*-}$
        & $45^{+0}_{-0}{}^{+58}_{-36}$
        & $45^{+0}_{-0}{}^{+35}_{-31}$
        & $44^{+0}_{-0}{}^{+23}_{-23}$
        & $50\pm 7$
        \\
      & $\pphi\Kbar^{*0}$
        & $44^{+0}_{-0}{}^{+59}_{-36}$
        & $44^{+0}_{-0}{}^{+35}_{-31}$
        & $43^{+0}_{-0}{}^{+23}_{-23}$
        & $49\pm 3$
        \\
        \addlinespace
    $A_{\CP}^0 / \%$
      & $\pphi\K^{*-}$
        & $-1^{+0}_{-0}{}^{+2}_{-1}$
        & $-1^{+0}_{-0}{}^{+1}_{-1}$
        & $-1^{+0}_{-0}{}^{+2}_{-2}$
        & n/a
        \\
      & $\pphi\Kbar^{*0}$
        & $\phantom{-}0^{+0}_{-0}{}^{+1}_{-1}$
        & $\phantom{-}0^{+0}_{-0}{}^{+1}_{-0}$
        & $\phantom{-}0^{+0}_{-0}{}^{+1}_{-2}$
        & $2\pm 7$
        \\
        \addlinespace
    $(f_\parallel-f_\perp) / \%$
      & $\pphi\K^{*-}$
        & $0^{+0}_{-0}{}^{+2}_{-2}$
        & $0^{+0}_{-0}{}^{+2}_{-2}$
        & $0^{+0}_{-0}{}^{+2}_{-2}$
        & $12\pm 17$
        \\
      & $\pphi\Kbar^{*0}$
        & $0^{+0}_{-0}{}^{+2}_{-2}$
        & $0^{+0}_{-0}{}^{+2}_{-2}$
        & $0^{+0}_{-0}{}^{+2}_{-2}$
        & $1\pm 7$
        \\
        \addlinespace
    $(A_{\CP}^\parallel-A_{\CP}^\perp) / \%$
      & $\pphi\K^{*-}$
        & $0^{+0}_{-0}{}^{+0}_{-0}$
        & $0^{+0}_{-0}{}^{+0}_{-0}$
        & $0^{+0}_{-0}{}^{+0}_{-0}$
        & n/a
        \\
      & $\pphi\Kbar^{*0}$
        & $0^{+0}_{-0}{}^{+0}_{-0}$
        & $0^{+0}_{-0}{}^{+0}_{-0}$
        & $0^{+0}_{-0}{}^{+0}_{-0}$
        & $18\pm 28$
        \\
        \addlinespace
    $\phi_\parallel/^\circ$
      & $\pphi\K^{*-}$
        & $-41^{+0}_{-0}{}^{+84}_{-53}$
        & $-41^{+0}_{-0}{}^{+35}_{-30}$
        & $-40^{+0}_{-0}{}^{+21}_{-21}$
        & $-60\pm 16$
        \\
      & $\pphi\Kbar^{*0}$
        & $-42^{+0}_{-0}{}^{+87}_{-54}$
        & $-42^{+0}_{-0}{}^{+35}_{-30}$
        & $-42^{+0}_{-0}{}^{+21}_{-21}$
        & $-44\pm 8$
        \\
        \addlinespace
    $\Delta\phi_\parallel/^\circ$
      & $\pphi\K^{*-}$
        & $0^{+0}_{-0}{}^{+0}_{-1}$
        & $0^{+0}_{-0}{}^{+0}_{-0}$
        & $0^{+0}_{-0}{}^{+0}_{-0}$
        & n/a
        \\
      & $\pphi\Kbar^{*0}$
        & $0^{+0}_{-0}{}^{+0}_{-0}$
        & $0^{+0}_{-0}{}^{+0}_{-0}$
        & $0^{+0}_{-0}{}^{+0}_{-1}$
        & $6\pm 8$
        \\
        \addlinespace
    $(\phi_\parallel-\phi_\perp)/^\circ$
      & $\pphi\K^{*-}$
        & $0^{+0}_{-0}{}^{+1}_{-1}$
        & $0^{+0}_{-0}{}^{+1}_{-1}$
        & $0^{+0}_{-0}{}^{+1}_{-1}$
        & $-12\pm 24$
        \\
      & $\pphi\Kbar^{*0}$
        & $0^{+0}_{-0}{}^{+1}_{-1}$
        & $0^{+0}_{-0}{}^{+1}_{-1}$
        & $0^{+0}_{-0}{}^{+1}_{-1}$
        & $1\pm 11$
        \\
        \addlinespace
    $(\Delta\phi_\parallel-\Delta\phi_\perp)/^\circ$
      & $\pphi\K^{*-}$
        & $0^{+0}_{-0}{}^{+0}_{-0}$
        & $0^{+0}_{-0}{}^{+0}_{-0}$
        & $0^{+0}_{-0}{}^{+0}_{-0}$
        & n/a
        \\
      & $\pphi\Kbar^{*0}$
        & $0^{+0}_{-0}{}^{+0}_{-0}$
        & $0^{+0}_{-0}{}^{+0}_{-0}$
        & $0^{+0}_{-0}{}^{+0}_{-0}$
        & $3\pm 11$
        \\
    \bottomrule
  \end{tabular}
  \let\arraystretch=\oldarraystretch
  \caption{Comparison of theoretical results for observables from the
    full angular analysis of $\Bm\to\pphi\K^{*-}$ and 
    $\Bdq\to\pphi\bar{\K}^{*0}$ with experimental results 
    from~\cite{Aubert:2004xc,newbabar,Chen:2005zv}.}
  \label{tab:phikst}
\end{table}

\begin{figure}
  \includegraphics[width=0.3\linewidth]{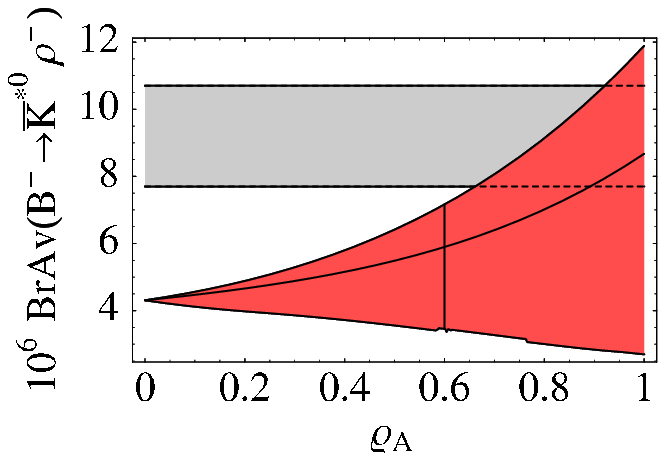}
  \hfill
  \includegraphics[width=0.3\linewidth]{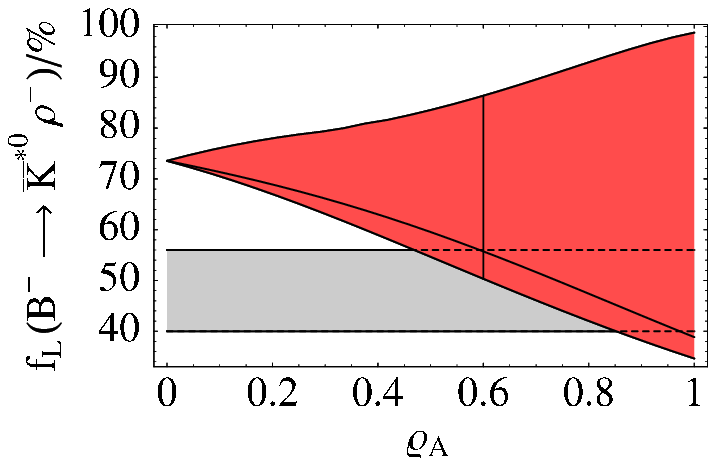}
  \hfill
  \includegraphics[width=0.3\linewidth]{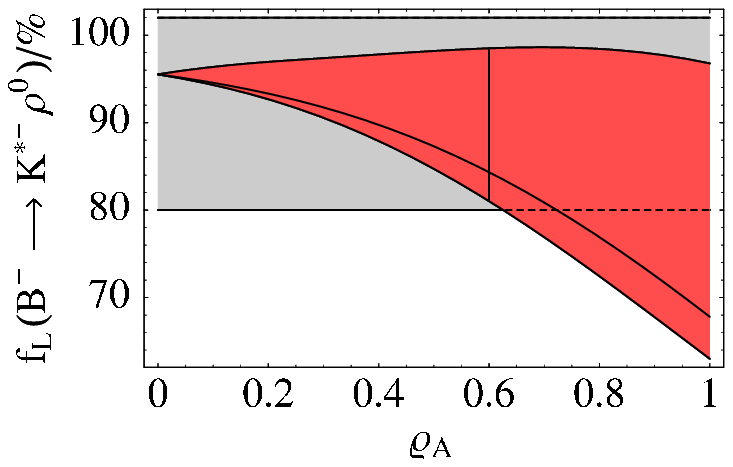}
  \caption{Predicted branching fraction and longitudinal polarisation
    fraction of the pure penguin decay $\Bm\to\Kbar^{*0}\prho^-$ (left
    and center) and longitudinal polarisation fraction of
    $\Bm\to\K^{*-}\prho^0$ (right) as a function of the parameter
    $\varrho_A$ of our annihilation model, showing only the theory
    uncertainty from variation of the phase $\varphi_A$. The
    horizontal band represents current experimental
    values~\cite{hfag2006}. The lines within the theory band indicate
    our default choices for $\varrho_A$ and $\varphi_A$.}
  \label{fig:rhokst_BrAv-XA}
\end{figure}

We now explore a strategy where the variation of input parameters or
the transverse penguin amplitude $\hat\alpha_4^{p-}$ is constrained by
data in order to improve the predictions for other observables and
decay modes. We assume that this amplitude is approximately the same
for all decay modes, in accordance with factorisation calculations.
Figure~\ref{fig:kstphi_transAmp} allows explanations for the large
negative-helicity amplitude based on spectator scattering or weak
annihilation (or both). We favour the second option, since an
enhancement of spectator scattering would lead to large transverse
polarisation for the colour-suppressed tree decays which is not
observed for the $\rho^0\rho^0$ final state, see
Section~\ref{sec:tree}. Further support for this option comes from the
$\rho\K^*$ final states, which do not involve $\alpha_3^p$, and which
are therefore much less sensitive to spectator scattering.
Figure~\ref{fig:rhokst_BrAv-XA} displays three observables from the
$\rho\K^*$ system as a function of the strength of weak annihilation,
$\varrho_A$. It can be seen that it must be non-zero, with a favoured
range around $\varrho_A\approx 0.6$. This is consistent with a fit of
$\varrho_A \,e^{i\varphi_A}$ to the $\pphi\K^{*0}$ data, which
suggests
\begin{equation}
  \varrho_A = 0.5 \pm 0.2_\text{exp.} 
  \qquad
  \varphi_A = (-43 \pm 19_\text{exp.})^\circ,
  \label{fitrhoA}
\end{equation}
excluding other theory uncertainties. This coincidence motivates the
default input parameter choice $\varrho_A \,e^{i\varphi_A}
=0.6\,e^{-i40^\circ}$ adopted in Section~\ref{sec:inputs}. The second
column in Table~\ref{tab:phikst} shows the theoretical prediction when
the variation of these parameters is reduced to $\varrho_A=0.6\pm0.2$
and $\varphi_A=(-40\pm 10)^\circ$ as suggested by (\ref{fitrhoA}) and
Figure~\ref{fig:rhokst_BrAv-XA}. The central values of these results
remain the same by construction, but the hadronic uncertainties are
considerably reduced.

Rather than relying on our model-dependent parameterisation of the
weak annihilation amplitude to fit data, we prefer the point of view
that $\hat{\alpha}_4^{p-}$ is theoretically unreliable and should be
taken from data. Thus, instead of $\varrho_A \,e^{i\varphi_A}$ we fit
$\hat{\alpha}_4^{p-}$. Neglecting CP violation in the $\phi\K^*$
system in accordance with the present data and theoretical
expectations, the transverse amplitude can be expressed as
\begin{equation}
  \bar{\Amp}_- = A_{\Kbar^*\phi} \lambda^{(s)}_c P^{\Kbar^*\phi}_-,
\end{equation}
where, neglecting small coefficients, $P^{\Kbar^*\phi}_- \approx
\hat{\alpha}_4^{c-}(\Kbar^*\phi)+\alpha_3^c(\Kbar^*\phi)$. Calculating
$\bar{\Amp}_-$ from the data with the overall phase adapted so that
the phases of the longitudinal helicity amplitudes from theory and
data match, we obtain
\begin{equation}
  P^{\Kbar^*\phi}_- =
          (-0.081 \pm 0.002\text{(exp)}{}^{+0.008}_{-0.009}\text{(th)})
    + \I\,(0.026 \pm 0.011\text{(exp)}{}^{+0.003}_{-0.003}\text{(th)}),
  \label{eq:Pkstphi}
\end{equation}
where the theoretical error includes the uncertainties from $A_{\Kbar^*\phi}
\lambda^{(s)}_c$ only. Since we already assumed that a large 
enhancement of $\alpha_3^{c-}$ is not a favoured option, we 
obtain  $\hat{\alpha}_4^{c-}$ by subtracting the default value 
of $\alpha_3^{c-}$. Rounding numbers, this results in 
\begin{equation}
  \hat{\alpha}_4^{c-} = (-0.08\pm0.02) + \I\,(0.03 \pm 0.02). 
\label{al4hat}
\end{equation}
This is the input to the theoretical predictions shown in the third 
column of Table~\ref{tab:phikst} and the columns labeled 
``$\hat{\alpha}_4^{c-}$ from data'' in later tables in this section. 
This procedure provides another considerable
reduction of hadronic uncertainties. 

\subsection{Branching fractions, direct CP asymmetries and
  polarisation}

We now discuss the complete set of final states, where the
$\hat{\alpha}_4^c$ amplitude is dominant. The CP-averaged 
branching fractions, direct CP asymmetries and longitudinal 
polarisation fractions are given in 
Tables~\ref{tab:penguinBrAv},~\ref{tab:penguinACP}
and~\ref{tab:penguinfL}, respectively. The two columns of numbers 
in these tables 
represent the result with our default inputs with uncertainties, and 
the result using (\ref{al4hat}) as input. In general, we regard 
the second result as our ``best'' prediction. However, one 
should be aware that it depends on the assumption 
of final-state independence of $\hat{\alpha}_4^c$, and the 
present experimental data. 

\begin{table}
  \centering
  \let\oldarraystretch=\arraystretch
  \renewcommand*{\arraystretch}{1.2}
  \begin{tabular}{llll}
    \toprule
        $\BrAv / 10^{-6}$
        & \multicolumn{2}{l}{Theory}
        & Experiment
        \\
    \cmidrule(l){2-3}
        & default
        & $\hat{\alpha}_4^{c-}$ from data
        &
        \\
    \midrule
    $\Bm\to\K^{*-} \pphi$
        & $10.1^{+0.5}_{-0.5}{}^{+12.2}_{-7.1}$
        & $10.4^{+0.5}_{-0.5}{}^{+5.2}_{-3.9}$
        & $9.7\pm 1.5$
        \\
    $\Bdq\to\Kbar^{*0} \pphi$
        & $9.3^{+0.5}_{-0.5}{}^{+11.4}_{-6.5}$
        & $9.6^{+0.5}_{-0.5}{}^{+4.7}_{-3.6}$
        & $9.5\pm 0.8$
        \\
    $\Bm\to\K^{*-} \pomega$
        & $2.4^{+0.8}_{-0.7}{}^{+2.9}_{-1.3}$
        & $2.3^{+0.8}_{-0.7}{}^{+1.4}_{-0.7}$
        & $< 3.4$
        \\
    $\Bdq\to\Kbar^{*0} \pomega$
        & $2.0^{+0.1}_{-0.1}{}^{+3.1}_{-1.4}$
        & $1.9^{+0.1}_{-0.1}{}^{+1.5}_{-0.7}$
        & $< 4.2$
        \\
    $\Bm\to\Kbar^{*0} \prho^-$
        & $5.9^{+0.3}_{-0.3}{}^{+6.9}_{-3.7}$
        & $5.8^{+0.3}_{-0.3}{}^{+3.1}_{-1.9}$
        & $9.2\pm 1.5$
        \\
    $\Bm\to\K^{*-} \prho^0$
        & $4.5^{+1.5}_{-1.3}{}^{+3.0}_{-1.4}$
        & $4.5^{+1.5}_{-1.3}{}^{+1.8}_{-1.0}$
        & $< 6.1$
        \\
    $\Bdq\to\K^{*-} \prho^+$
        & $5.5^{+1.7}_{-1.5}{}^{+5.7}_{-2.9}$
        & $5.4^{+1.7}_{-1.5}{}^{+2.6}_{-1.5}$
        & $<12$
        \\
    $\Bdq\to\Kbar^{*0} \prho^0$
        & $2.4^{+0.2}_{-0.1}{}^{+3.5}_{-2.0}$
        & $2.3^{+0.2}_{-0.1}{}^{+1.1}_{-0.8}$
        & $5.6\pm 1.6$
        \\\addlinespace
    $\Bsq\to\K^{*-} \K^{*+}$
        & $9.1^{+2.5}_{-2.2}{}^{+10.2}_{-5.9}$
        & $8.0^{+2.4}_{-2.1}{}^{+3.7}_{-3.4}$
        & n/a
        \\
    $\Bsq\to\K^{*0} \Kbar^{*0}$
        & $9.1^{+0.5}_{-0.4}{}^{+11.3}_{-6.8}$
        & $7.9^{+0.4}_{-0.4}{}^{+4.3}_{-3.9}$
        & n/a
        \\
    $\Bsq\to\pphi \pphi$
        & $21.8^{+1.1}_{-1.1}{}^{+30.4}_{-17.0}$
        & $19.5^{+1.0}_{-1.0}{}^{+13.1}_{-8.0}$
        & $14.0^{+8.0}_{-7.0}$
        \\\addlinespace[2\defaultaddspace]
    $\Bm\to\K^{*0} \K^{*-}$
        & $0.5^{+0.2}_{-0.1}{}^{+0.4}_{-0.3}$
        & $0.5^{+0.2}_{-0.1}{}^{+0.2}_{-0.2}$
        & $<71$
        \\
    $\Bdq\to\K^{*0} \Kbar^{*0}$
        & $0.6^{+0.1}_{-0.1}{}^{+0.5}_{-0.3}$
        & $0.6^{+0.1}_{-0.1}{}^{+0.3}_{-0.2}$
        & $<22$
        \\\addlinespace
    $\Bsq\to\K^{*0} \pphi$
        & $0.4^{+0.1}_{-0.1}{}^{+0.5}_{-0.3}$
        & $0.3^{+0.1}_{-0.1}{}^{+0.2}_{-0.1}$
        & n/a
        \\
    \bottomrule
  \end{tabular}
  \let\arraystretch=\oldarraystretch
  \caption{CP-averaged branching fractions of $\Delta S=1$ and $\Delta
    D=1$ penguin-dominated $B\to VV$ decays. Experimental values
    are taken
    from~\cite{Aubert:2004xc,newbabar,Chen:2005zv,Aubert:2006vt,Aubert:2006fs,Abe:2004mq,Godang:2001sg}.}
  \label{tab:penguinBrAv}
\end{table}

The two $\pomega\K^*$ modes are predicted to have 4--5 times smaller 
branching fractions than the $\pphi\K^*$ modes, which is consistent 
with experimental upper limits. Following the notation of the appendix 
of~\cite{BN2003} and suppressing helicity labels $h$, the decay 
amplitudes read
\begin{equation}
  \begin{aligned}
    \sqrt{2} \Amp_{\Bm\to\pomega\K^{*-}} &= 
        A_{\Kbar^*\pomega} \left[2\alpha_3^p+\delta_{pu}\,\alpha_2\right]
      + A_{\pomega\Kbar^*} \left[\hat{\alpha}_4^p+\delta_{pu}\,\alpha_1\right],
    \\[0.2cm]
    \sqrt{2} \Amp_{\Bdq\to\pomega\Kbar^{*0}} &=
        A_{\Kbar^*\pomega} \left[2\alpha_3^p+\delta_{pu}\,\alpha_2\right]
      + A_{\pomega\Kbar^*}       \hat{\alpha}_4^p,
  \end{aligned}
\label{omkamps}
\end{equation}
where we do not show numerically irrelevant amplitudes. The smaller
branching fraction is simply a consequence of a relative factor of
$\sqrt{2}$ in the amplitude, and the smaller $B\to\omega$ form factors
multiplying the dominant $\hat{\alpha}_4^c$ coefficient
($|A_{\Kbar^{*}\pphi}^{0,-}/A_{\omega\Kbar^{*}}^{0,-}|^2 \approx
2.5$).  As both factors cancel out in the polarisation fraction, $f_L$
is predicted to be similar for $\omega K^*$ and $\pphi\K^*$. However,
contrary to the $\phi\K^*$ system, the presence of tree amplitudes
$\alpha_{1,2}$ in (\ref{omkamps}) allows for sizeable CP asymmetries
as can be seen from Table~\ref{tab:penguinACP}.

\begin{table}
  \centering
  \let\oldarraystretch=\arraystretch
  \renewcommand*{\arraystretch}{1.2}
  \begin{tabular}{llll}
    \toprule
        $A_{\CP}$ / percent
        & \multicolumn{2}{l}{Theory}
        & Experiment
        \\
    \cmidrule(l){2-3}
        & default
        & $\hat{\alpha}_4^{c-}$ from data
        &
        \\
    \midrule
    $\Bm\to\K^{*-} \pphi$
        & $\phantom{1}0^{+0}_{-0}{}^{+2}_{-1}$
        & $\phantom{1}0^{+0}_{-0}{}^{+3}_{-2}$
        & $5\pm 11$
        \\
    $\Bdq\to\Kbar^{*0} \pphi$
        & $\phantom{1}1^{+0}_{-0}{}^{+1}_{-0}$
        & $\phantom{1}1^{+0}_{-0}{}^{+2}_{-1}$
        & $1\pm 6$
        \\
    $\Bm\to\K^{*-} \pomega$
        & $22^{+4}_{-5}{}^{+40}_{-21}$
        & $24^{+5}_{-6}{}^{+25}_{-24}$
        & n/a
        \\
    $\Bdq\to\Kbar^{*0} \pomega$
        & $19^{+5}_{-4}{}^{+17}_{-16}$
        & $20^{+6}_{-5}{}^{+16}_{-15}$
        & n/a
        \\
    $\Bm\to\Kbar^{*0} \prho^-$
        & $\phantom{1}0^{+0}_{-0}{}^{+3}_{-1}$
        & $\phantom{1}0^{+0}_{-0}{}^{+3}_{-2}$
        & $-1\pm 16$
        \\
    $\Bm\to\K^{*-} \prho^0$
        & $16^{+4}_{-4}{}^{+23}_{-16}$
        & $16^{+4}_{-4}{}^{+17}_{-14}$
        & $20^{+32}_{-29}$
        \\
    $\Bdq\to\K^{*-} \prho^+$
        & $\phantom{1}5^{+1}_{-1}{}^{+40}_{-17}$
        & $\phantom{1}6^{+2}_{-2}{}^{+16}_{-12}$
        & n/a
        \\
    $\Bdq\to\Kbar^{*0} \prho^0$
        & $-15^{+4}_{-4}{}^{+17}_{-32}$
        & $-15^{+4}_{-4}{}^{+16}_{-16}$
        & $9\pm 19$
        \\\addlinespace
    $\Bsq\to\K^{*-} \K^{*+}$
        & $\phantom{1}2^{+0}_{-0}{}^{+40}_{-15}$
        & $\phantom{1}4^{+1}_{-1}{}^{+20}_{-12}$
        & n/a
        \\
    $\Bsq\to\K^{*0} \Kbar^{*0}$
        & $\phantom{1}1^{+0}_{-0}{}^{+1}_{-0}$
        & $\phantom{1}1^{+0}_{-0}{}^{+2}_{-1}$
        & n/a
        \\
    $\Bsq\to\pphi \pphi$
        & $\phantom{1}1^{+0}_{-0}{}^{+1}_{-0}$
        & $\phantom{1}1^{+0}_{-0}{}^{+2}_{-1}$
        & n/a
        \\\addlinespace[2\defaultaddspace]
    $\Bm\to\K^{*0} \K^{*-}$
        & $\phantom{1}0^{+0}_{-0}{}^{+17}_{-40}$
        & $\phantom{1}1^{+0}_{-0}{}^{+30}_{-55}$
        & n/a
        \\
    $\Bdq\to\K^{*0} \Kbar^{*0}$
        & $-13^{+3}_{-4}{}^{+6}_{-8}$
        & $-13^{+3}_{-4}{}^{+17}_{-24}$
        & n/a
        \\\addlinespace
    $\Bsq\to\K^{*0} \pphi$
        & $-17^{+4}_{-5}{}^{+9}_{-9}$
        & $-16^{+4}_{-4}{}^{+31}_{-51}$
        & n/a
        \\
    \bottomrule
  \end{tabular}
  \let\arraystretch=\oldarraystretch
  \caption{Direct CP asymmetries of $\Delta S=1$ and $\Delta
    D=1$ penguin-dominated $B\to VV$ decays. Experimental values
    are taken
    from~\cite{Aubert:2004xc,newbabar,Chen:2005zv,Aubert:2006vt,Aubert:2006fs}.}
  \label{tab:penguinACP}
\end{table}

\begin{table}[t]
  \centering
  \let\oldarraystretch=\arraystretch
  \renewcommand*{\arraystretch}{1.2}
  \begin{tabular}{llll}
    \toprule
        $f_L$ / percent
        & \multicolumn{2}{l}{Theory}
        & Experiment
        \\
    \cmidrule(l){2-3}
        & default
        & $\hat{\alpha}_4^{c-}$ from data
        &
        \\
    \midrule
    $\Bm\to\K^{*-} \pphi$
        & $45^{+0}_{-0}{}^{+58}_{-36}$
        & $44^{+0}_{-0}{}^{+23}_{-23}$
        & $50.0\pm 7.0$
        \\
    $\Bdq\to\Kbar^{*0} \pphi$
        & $44^{+0}_{-0}{}^{+59}_{-36}$
        & $43^{+0}_{-0}{}^{+23}_{-23}$
        & $49.1\pm 3.2$
        \\
    $\Bm\to\K^{*-} \pomega$
        & $53^{+8}_{-11}{}^{+57}_{-39}$
        & $56^{+8}_{-11}{}^{+22}_{-19}$
        & n/a
        \\
    $\Bdq\to\Kbar^{*0} \pomega$
        & $40^{+4}_{-3}{}^{+77}_{-43}$
        & $43^{+4}_{-3}{}^{+38}_{-32}$
        & n/a
        \\
    $\Bm\to\Kbar^{*0} \prho^-$
        & $56^{+0}_{-0}{}^{+48}_{-30}$
        & $57^{+0}_{-0}{}^{+21}_{-18}$
        & $48.0\pm 8.0$
        \\
    $\Bm\to\K^{*-} \prho^0$
        & $84^{+2}_{-3}{}^{+16}_{-25}$
        & $85^{+2}_{-3}{}^{+9}_{-11}$
        & $96^{+6}_{-16}$
        \\
    $\Bdq\to\K^{*-} \prho^+$
        & $61^{+5}_{-7}{}^{+38}_{-28}$
        & $62^{+5}_{-6}{}^{+17}_{-15}$
        & n/a
        \\
    $\Bdq\to\Kbar^{*0} \prho^0$
        & $22^{+3}_{-3}{}^{+53}_{-14}$
        & $22^{+3}_{-3}{}^{+21}_{-13}$
        & $57\pm 12$
        \\\addlinespace
    $\Bsq\to\K^{*-} \K^{*+}$
        & $67^{+4}_{-5}{}^{+31}_{-26}$
        & $76^{+3}_{-4}{}^{+12}_{-16}$
        & n/a
        \\
    $\Bsq\to\K^{*0} \Kbar^{*0}$
        & $63^{+0}_{-0}{}^{+42}_{-29}$
        & $72^{+0}_{-0}{}^{+16}_{-21}$
        & n/a
        \\
    $\Bsq\to\pphi \pphi$
        & $43^{+0}_{-0}{}^{+61}_{-34}$
        & $48^{+0}_{-0}{}^{+26}_{-27}$
        & n/a
        \\\addlinespace[2\defaultaddspace]
    $\Bm\to\K^{*0} \K^{*-}$
        & $62^{+1}_{-2}{}^{+42}_{-33}$
        & $62^{+1}_{-2}{}^{+18}_{-19}$
        & n/a
        \\
    $\Bdq\to\K^{*0} \Kbar^{*0}$
        & $69^{+1}_{-1}{}^{+34}_{-27}$
        & $69^{+1}_{-1}{}^{+16}_{-20}$
        & n/a
        \\\addlinespace
    $\Bsq\to\K^{*0} \pphi$
        & $40^{+1}_{-1}{}^{+67}_{-35}$
        & $47^{+3}_{-2}{}^{+28}_{-24}$
        & n/a
        \\
    \bottomrule
  \end{tabular}
  \let\arraystretch=\oldarraystretch
  \caption{Longitudinal polarisation fraction of $\Delta S=1$ and $\Delta
    D=1$ penguin-dominated $B\to VV$ decays. Experimental values
    are taken from~\cite{Aubert:2004xc,newbabar,Chen:2005zv,Aubert:2006vt,Aubert:2006fs,Abe:2004mq}.}
  \label{tab:penguinfL}
\end{table}

The four $\K^*\prho$ final states are the $VV$ equivalents to the 
much discussed $\pi K$ final states. Their amplitudes are given by 
\begin{equation}
  \begin{aligned}
    \Amp_{\Bm\to\prho^-\Kbar^{*0}} &=
      A_{\rho\Kbar^*} \,\hat{\alpha}_4^p,
    \\
    \sqrt{2}\,\Amp_{\Bm\to\prho^0\K^{*-}} &=
      A_{\rho\Kbar^*} \big[ \hat{\alpha}_4^p + \delta_{pu}\alpha_1 \big]
    + A_{\Kbar^*\rho} \left[ \frac{3}{2} \alpha_{3,\EW}^p 
                           +\delta_{pu}\alpha_2 \right],
    \\
    \Amp_{\Bdq\to\prho^+\K^{*-}} &=
      A_{\rho\Kbar^*} \big[ \hat{\alpha}_4^p + \delta_{pu}\alpha_1 \big],
    \\
    -\sqrt{2}\,\Amp_{\Bdq\to\prho^0\Kbar^{*0}} &=
      A_{\rho\Kbar^*} \, \hat{\alpha}_4^p
    - A_{\Kbar^*\rho} \left[ \frac{3}{2} \alpha_{3,\EW}^p 
                           +\delta_{pu}\alpha_2 \right]
    .
  \end{aligned}
  \label{eq:rhokst}
\end{equation}
They are particularly interesting, because the colour-allowed
electroweak penguin amplitude $\alpha_{3,\EW}^{c-}$ plays an important
role, in particular for polarisation. We discuss this point separately
in Section~\ref{subsec:rhokst}. Here we note with respect to
Table~\ref{tab:penguinBrAv} that the branching fractions seem to be
systematically below the measurements. Since the $\Bm\to \bar
K^{*0}\rho^-$ mode is a pure penguin decay, proportional to the
$B\to\rho$ form factor, this is problematic. A larger $B\to\rho$ form
factor is not an option, since this would be in conflict with the
observed $B\to\rho\rho$ branching fractions. A larger value of
$\hat{\alpha}_4^{c,0}$ would also increase the $\phi K^*$ branching
fractions, unless the penguin amplitude is highly non-universal, or
the $B\to K^*$ form factors are smaller than assumed, or one arranges
a cancellation between $\alpha_4^{c,h}$ and $\alpha_3^{c,h}$ in $\phi
K^*$. Rather than pursuing any of these options, we leave this issue
as a potential problem for the QCD factorisation approach (or the
input parameter set). The magnitude and sign of direct CP asymmetries
is again related to the presence or absence of tree amplitudes that
can interfere with the leading QCD penguin. Therefore, we predict
negligible $A_{\CP}$ for the pure-penguin mode $\bar K^{*0}\prho^-$,
while asymmetries up to about $30\%$ are possible for some of the
other modes.

The list of $\Delta S = 1$ penguin-dominated decays terminates with
three $\Bsq$ modes. Here the flavour topology allows the
penguin-annihilation amplitude $\beta_4^c$, which turns out to be the
first subdominant contribution besides $\hat{\alpha}_4^c$ for
the longitudinal amplitude, but is negligible for the transverse ones.
Including these coefficients, the simplified amplitude expressions
read
\begin{equation}
  \begin{aligned}
    \Amp_{\Bsq\to\K^{*0}\Kbar^{*0}} &=
      A_{\Kbar^*\K^*} \beta_4^p
    + A_{\K^*\Kbar^*} \left[ \hat{\alpha}_4^p + \beta_4^p \right],
    \\[0.2cm]
    \Amp_{\Bsq\to\K^{*-}\K^{*+}} &=
      A_{\Kbar^*\K^*} \beta_4^p
    + A_{\K^*\Kbar^*} \left[
          \hat{\alpha}_4^p + \beta_4^p + \delta_{pu}\alpha_1
        \right],
  \end{aligned}
\end{equation}
\begin{equation*}
    \frac{1}{2} \Amp_{\Bsq\to\pphi\pphi} =
      A_{\pphi\pphi} \left[
         \hat{\alpha}_4^p + \alpha_3^p - \frac{1}{2} \alpha_{3\EW}^p 
         + \beta_4^p \right].
\end{equation*}
The relative factor of two in the $\pphi\pphi$ amplitude compared to
the others leads to a particularly large branching fraction 
for this decay, even though the enhancement is somewhat reduced due 
to the destructive interference of the electroweak penguin and 
colour-suppressed QCD penguin amplitude in the longitudinal amplitude, 
see Table~\ref{tab:ampratios}. The interference is constructive 
for the negative-helicity amplitude, leading to smaller $f_L$ 
for $\bar B_s\to\phi\phi$ compared to the other two $\Bsq$
decays. Sizeable CP asymmetries are only expected for the
decay $\Bsq\to\K^{*-}\K^{*+}$ with a tree contribution.

We briefly examine the three $\Delta D=1$
penguin channels, which have small branching fractions. The first
subdominant contribution in these decays comes from the
$\hat{\alpha}_4^u$ up-penguin amplitude. In theoretical calculations 
the strong phases of $\hat{\alpha}_4^c$ and $\hat{\alpha}_4^u$ are correlated 
to a certain extent. This information is lost for the 
transverse amplitude when $\hat{\alpha}_4^{c-}$ is taken from 
data and  $\hat{\alpha}_4^{u-}$ is calculated. This explains why the 
error on the CP asymmetries increases  for these decays from the first to 
the second columns of numbers in Table~\ref{tab:penguinACP}.

\begin{table}[t]
  \centering
  \let\oldarraystretch=\arraystretch
  \renewcommand*{\arraystretch}{1.2}
  \begin{tabular}{lllllll}
    \toprule
        & \multicolumn{2}{l}{$\phi_\parallel / ^\circ$}
        & \multicolumn{2}{l}{$A_{\CP}^0$ / percent}
        & \multicolumn{2}{l}{$\Delta\phi_\parallel / ^\circ$}
        \\
    \cmidrule(l){2-3}
    \cmidrule(l){4-5}
    \cmidrule(l){6-7}
        & default & $\hat{\alpha}_4^{c-}$ f.\,d.
        & default & $\hat{\alpha}_4^{c-}$ f.\,d.
        & default & $\hat{\alpha}_4^{c-}$ f.\,d.
        \\
    \midrule
    $\Bm\to\K^{*-} \pphi$
        & $-41^{+0}_{-0}{}^{+84}_{-53}$
        & $-40^{+0}_{-0}{}^{+21}_{-21}$
        & $-1^{+0}_{-0}{}^{+2}_{-1}$
        & $-1^{+0}_{-0}{}^{+2}_{-2}$
        & $0^{+0}_{-0}{}^{+0}_{-1}$
        & $0^{+0}_{-0}{}^{+1}_{-1}$
        \\
    $\Bdq\to\Kbar^{*0} \pphi$
        & $-42^{+0}_{-0}{}^{+87}_{-54}$
        & $-42^{+0}_{-0}{}^{+21}_{-21}$
        & $0^{+0}_{-0}{}^{+1}_{-1}$
        & $0^{+0}_{-0}{}^{+1}_{-2}$
        & $0^{+0}_{-0}{}^{+0}_{-0}$
        & $0^{+0}_{-0}{}^{+1}_{-1}$
        \\
    $\Bm\to\K^{*-} \pomega$
        & $-33^{+5}_{-6}{}^{+113}_{-72}$
        & $-31^{+5}_{-6}{}^{+18}_{-20}$
        & $28^{+17}_{-10}{}^{+56}_{-57}$
        & $26^{+17}_{-10}{}^{+45}_{-40}$
        & $-32^{+8}_{-9}{}^{+73}_{-34}$
        & $-31^{+8}_{-9}{}^{+26}_{-30}$
        \\
    $\Bdq\to\Kbar^{*0} \pomega$
        & $-43^{+3}_{-3}{}^{+106}_{-73}$
        & $-41^{+3}_{-3}{}^{+35}_{-38}$
        & $15^{+5}_{-4}{}^{+61}_{-51}$
        & $14^{+5}_{-4}{}^{+52}_{-50}$
        & $-11^{+3}_{-4}{}^{+23}_{-32}$
        & $-12^{+3}_{-4}{}^{+20}_{-29}$
        \\
    $\Bm\to\Kbar^{*0} \prho^-$
        & $-37^{+0}_{-0}{}^{+92}_{-59}$
        & $-35^{+0}_{-0}{}^{+18}_{-19}$
        & $-1^{+0}_{-0}{}^{+1}_{-1}$
        & $0^{+0}_{-0}{}^{+1}_{-1}$
        & $0^{+0}_{-0}{}^{+0}_{-2}$
        & $0^{+0}_{-0}{}^{+1}_{-1}$
        \\
    $\Bm\to\K^{*-} \prho^0$
        & $-39^{+4}_{-5}{}^{+146}_{-88}$
        & $-37^{+4}_{-5}{}^{+21}_{-20}$
        & $7^{+2}_{-2}{}^{+12}_{-13}$
        & $6^{+2}_{-2}{}^{+8}_{-8}$
        & $-14^{+3}_{-4}{}^{+29}_{-60}$
        & $-13^{+3}_{-4}{}^{+15}_{-15}$
        \\
    $\Bdq\to\K^{*-} \prho^+$
        & $-36^{+4}_{-5}{}^{+111}_{-68}$
        & $-34^{+4}_{-5}{}^{+16}_{-16}$
        & $18^{+6}_{-5}{}^{+12}_{-29}$
        & $17^{+6}_{-5}{}^{+13}_{-12}$
        & $-19^{+5}_{-5}{}^{+74}_{-18}$
        & $-18^{+5}_{-5}{}^{+8}_{-8}$
        \\
    $\Bdq\to\Kbar^{*0} \prho^0$
        & $-41^{+4}_{-4}{}^{+63}_{-44}$
        & $-39^{+4}_{-4}{}^{+18}_{-21}$
        & $-30^{+11}_{-11}{}^{+60}_{-48}$
        & $-30^{+10}_{-11}{}^{+57}_{-49}$
        & $17^{+5}_{-5}{}^{+22}_{-24}$
        & $17^{+5}_{-5}{}^{+22}_{-24}$
        \\\addlinespace
    $\Bsq\to\K^{*-} \K^{*+}\!\!\!$
        & $-34^{+3}_{-4}{}^{+113}_{-70}$
        & $-29^{+3}_{-4}{}^{+17}_{-23}$
        & $11^{+3}_{-3}{}^{+7}_{-17}$
        & $8^{+2}_{-2}{}^{+13}_{-7}$
        & $-17^{+4}_{-5}{}^{+105}_{-19}$
        & $-14^{+3}_{-4}{}^{+10}_{-10}$
        \\
    $\Bsq\to\K^{*0} \Kbar^{*0}$
        & $-34^{+0}_{-0}{}^{+110}_{-62}$
        & $-29^{+0}_{-0}{}^{+19}_{-26}$
        & $0^{+0}_{-0}{}^{+0}_{-0}$
        & $0^{+0}_{-0}{}^{+1}_{-1}$
        & $0^{+0}_{-0}{}^{+10}_{-3}$
        & $0^{+0}_{-0}{}^{+2}_{-1}$
        \\
    $\Bsq\to\pphi \pphi$
        & $-39^{+0}_{-0}{}^{+86}_{-57}$
        & $-37^{+0}_{-0}{}^{+21}_{-24}$
        & $0^{+0}_{-0}{}^{+1}_{-0}$
        & $0^{+0}_{-0}{}^{+1}_{-2}$
        & $0^{+0}_{-0}{}^{+0}_{-1}$
        & $0^{+0}_{-0}{}^{+1}_{-1}$
        \\\addlinespace[2\defaultaddspace]
    $\Bm\to\K^{*0} \K^{*-}\!\!\!$
        & $-39^{+2}_{-3}{}^{+96}_{-57}$
        & $-38^{+2}_{-3}{}^{+18}_{-21}$
        & $9^{+3}_{-2}{}^{+12}_{-24}$
        & $8^{+2}_{-2}{}^{+24}_{-21}$
        & $-5^{+1}_{-1}{}^{+28}_{-7}$
        & $-5^{+1}_{-1}{}^{+12}_{-26}$
        \\
    $\Bdq\to\K^{*0} \Kbar^{*0}$
        & $-32^{+0}_{-0}{}^{+82}_{-51}$
        & $-31^{+0}_{-0}{}^{+18}_{-27}$
        & $0^{+0}_{-0}{}^{+2}_{-4}$
        & $-0^{+0}_{-0}{}^{+23}_{-16}$
        & $3^{+1}_{-1}{}^{+14}_{-6}$
        & $3^{+1}_{-1}{}^{+17}_{-26}$
        \\\addlinespace
    $\Bsq\to\K^{*0} \pphi$
        & $-49^{+2}_{-1}{}^{+110}_{-62}$
        & $-46^{+2}_{-2}{}^{+26}_{-24}$
        & $-9^{+2}_{-3}{}^{+16}_{-20}$
        & $-9^{+2}_{-3}{}^{+58}_{-31}$
        & $3^{+1}_{-1}{}^{+16}_{-6}$
        & $7^{+2}_{-2}{}^{+21}_{-30}$
        \\\addlinespace[1\defaultaddspace]
    \bottomrule
  \end{tabular}
  \let\arraystretch=\oldarraystretch
  \caption{Theoretical results for other polarisation observables in 
    penguin-dominated $\B\to VV$ decays.}
    \label{tab:penguinpol}
\end{table}

We conclude this subsection by providing the remaining polarisation
observables in Table~\ref{tab:penguinpol}. As we see from
Figure~\ref{fig:kstphi_transAmp}, the transverse phase observable
$\phi_\parallel$ is very sensitive on penguin weak annihilation, and
therefore uncertain, but it is expected to be nearly the same for all
14 decay modes. This is clearly also true when we fit $\alpha_4^{c-}$
to data, but in this case the errors are much smaller. For the
polarisation CP asymmetries, the same qualitative statements as for
the full direct CP asymmetries are valid, i.\,e.\ they can be
significantly different from zero only in decays with a tree
contribution or in the $\Delta D=1$ modes.  Only rough estimates for
these asymmetries can be given. We do not display the observables that
vanish when the positive-helicity amplitude is zero. We find that
$|f_\parallel-f_\perp| \leq 4\,\%$ for all modes, and that
$\phi_\parallel-\phi_\perp$ and
$\Delta\phi_\parallel-\Delta\phi_\perp$ never exceed $\pm 2^\circ$.
As discussed before, these statements should be taken with some
caution, since the positive-helicity amplitude is estimated in the
naive-factorisation approximation.

\subsection{\boldmath The $\B\to\prho\K^{*}$ system and the electromagnetic 
  penguin effect}
\label{subsec:rhokst}

The $\prho\K^{*}$ final states are particularly interesting for an 
investigation of electroweak penguin effects, since the suppression 
of the leading $VV$ QCD penguin amplitude makes the electroweak 
penguin amplitude sizeable in comparison (about 50\%). Moreover, 
as can be seen from (\ref{eq:rhokst}), the electroweak penguin 
enters the amplitudes in three different combinations,
\begin{equation}
A_{\rho\Kbar^*} \,\hat{\alpha}_4^{p,h} + 
k\cdot A_{\Kbar^*\rho} \,\frac{3}{2} \alpha_{3,\EW}^{p,h}, 
\qquad k=1,0,-1, 
\label{interferences}
\end{equation}
allowing various kinds of interferences. Another interesting point is 
that the pattern of interference is opposite for the longitudinal and 
negative-helicity amplitude, since the sign of $\alpha_{3,\EW}^{p,-}$ 
is different from $\alpha_{3,\EW}^{p,0}$. This comes from an additional, 
power-enhanced contribution to $\alpha_{3,\EW}^{p,-}$~\cite{BRYEWP}, 
such that 
\begin{equation} 
  {\alpha_{3,\EW}^{p,-}}_{|\rm excl.} 
  \to
  \alpha_{3,\EW}^{p,-} = 
  {\alpha_{3,\EW}^{p,-}}_{|\rm excl.} 
  - \frac{2\alpha_{\rm em}}{3\pi}
    C_{7\gamma}^{\rm eff}
    \frac{m_B m_b}{m_\rho^2},
  \label{extraal3ew}
\end{equation} 
which changes the real part from $-0.010^{+0.002}_{-0.002}$ to the value 
$+0.015^{+0.004}_{-0.003}$ given in Table~\ref{tab:ampratios}. Since the 
term proportional to the Wilson coefficient of the electromagnetic 
dipole operator, $C_{7\gamma}^{\rm eff}$, is the largest 
contribution to the negative-helicity electroweak penguin 
amplitude, the interference patterns (\ref{interferences}) are 
sensitive to possible anomalous contributions to $C_{7\gamma}^{\rm eff}$, 
including its phase. 

In Table~\ref{tab:rhokst} we compare selected 
observables for the two final states involving  
$\alpha_{3,\EW}^{p,h}$, when the extra term in (\ref{extraal3ew}) 
is excluded, to the default (included) and data. We note 
that already in the ``excluded'' results, the longitudinal 
polarisation fractions of the $\prho\K^{*}$ final states 
are predicted to differ such that $f_L(\K^{*-}\rho^0) 
> f_L(\bar K^{*0}\rho^-) > f_L(\bar K^{*0}\rho^0)$. This follows 
from the large longitudinal electroweak penguin contribution. 
The transverse electromagnetic dipole effect amplifies the 
hierarchy among the three $f_L$ predictions. The current 
experimental data confirm the first inequality, but the 
second is not seen.

\begin{table}
  \centering
  \let\oldarraystretch=\arraystretch
  \renewcommand*{\arraystretch}{1.2}
  \begin{tabular}{lrrrrrrc}
    \toprule
        & \multicolumn{3}{c}{$\Bm\to\K^{*-}\rho^0$}
        & \phantom{C}
        & \multicolumn{3}{c}{$\Bdq\to\Kbar^{*0}\rho^0$}\\
    \cmidrule(l){2-4}
    \cmidrule(l){6-8}
        & incl. & excl.& exp.
        &
        & incl. & excl.& exp. \\
    \midrule
    $\BrAv/10^{-6}$ 
       & $4.5$
       & $5.4$   
       & $<6.1$
       &    
       & $2.4$
       & $1.4$ 
       & $5.6\pm1.6$ \\
    $f_L$ / \%
       & $84$
       & $70$  
       & $96^{+6}_{-16}$     
        &    
      & $22$
       & $37$ 
      &$57\pm 12$\\
     $A_{\CP}$ / \%
       &$\phantom{-}16$
       &$ \phantom{-}14$      
       &$20^{+32}_{-29}$ 
        &    
      &$ -15$
       &$ -24$ 
      &$9\pm 19$ \\
     \bottomrule
  \end{tabular}
  \let\arraystretch=\oldarraystretch
  \caption{Predicted branching fraction, longitudinal polarisation and direct 
  CP asymmetry of the two $\rho K^*$ final states sensitive to the 
  electroweak penguin amplitude with the power-enhanced transverse 
  contribution proportional to $C_{7\gamma}$ included or excluded 
  (theoretical errors in earlier tables). 
  Experimental results for comparison (exp.).}
  \label{tab:rhokst}
\end{table}

Similar to what has been done for the $\pi K$ system 
in~\cite{BN2003,Neubert:1998pt,Buras:1998rb,Yoshikawa:2003hb,Gronau:2003kj,Buras:2003yc},
one can construct decay rate ratios that highlight the electroweak
penguin contribution. However, in contrast to $\pi K$, one cannot
expand in small amplitude ratios; the suppression of the $VV$ QCD
penguin amplitude makes these ratios too large.  In~\cite{BRYEWP} a
few amplitude ratios related to the transverse polarisation decay
rates have been discussed, and their dependence on the electromagnetic
dipole operator has been emphasized.  The default input of the present
analysis is similar to~\cite{BRYEWP}. The amplitude ratio $p_h^{\rm
  EW} = P^{\rm EW}_h/P_h$ used in~\cite{BRYEWP} is approximately equal
to $3\alpha_{3,\EW}^{c,h}/ (2 \hat\alpha_{4}^{c,h})$. We now calculate
$\mbox{Re}(p_-^{\rm EW}) =
-0.25^{+0.18}_{-1.12}\,[+0.15^{+1.03}_{-0.16}]$ 
compared to $\mbox{Re}(p_-^{\rm EW}) = -0.23 \pm
0.08\,[+0.14^{+0.04}_{-0.05}]$ in~\cite{BRYEWP}, where the numbers in
square brackets refer to the (unrealistic) scenario when the
electromagnetic dipole effect is switched off. Only a small part of
the large range of theoretically allowed values given in the first 
number is in fact
compatible with the observed branching fractions, so that the
determination of the longitudinal and negative-helicity 
$p^{\rm EW}_{0,-}$ from theory alone is 
not the optimal approach. Therefore, in~\cite{BRYEWP} the QCD penguin 
amplitudes $P_{0,-}$ have been obtained from the branching and
longitudinal polarisation fraction of the pure penguin decay 
$B^-\to \bar K^{*0}\rho^-$, and the assumption that the phase of 
$p^{\EW}_{0,-}$ does not exceed $\pm 30^\circ$ has been made. 
The experimental data that goes into this analysis has
changed since; most importantly, the current smaller value of
$f_L(\bar K^{*0}\rho^-)$ implies a larger transverse QCD penguin
amplitude.  Repeating the fit in~\cite{BRYEWP} with current data, 
we now obtain
$\mbox{Re}(p_-^{\rm EW}) = -0.21^{+0.07}_{-0.08}\,[+0.11 \pm 0.06]$, 
which reduces the impact of electroweak penguins by about
20\% relative to the calculation, but dramatically improves on 
the theoretical error. 

\begin{table}
  \centering
  \let\oldarraystretch=\arraystretch
  \renewcommand*{\arraystretch}{1.2}
  \begin{tabular}{lrrrrr}
    \toprule
        & \multicolumn{2}{c}{Default}
        & \phantom{C}
        & \multicolumn{2}{c}{$P_h$ from data}\\
    \midrule
    $S_0$ 
       & $0.36^{+0.06}_{-0.05}{}^{+0.19}_{-0.13}$
       & 
       &    
       & $0.40^{+0.07}_{-0.07}{}^{+0.17}_{-0.16}$
       &  \\
    $S^\prime_0 $
       & $2.34^{+0.89}_{-0.75}{}^{+0.72}_{-0.73}$
       &   
       & 
       & $2.10^{+0.74}_{-0.63}{}^{+0.23}_{-0.25}$
       & \\
    $S^{''}_0$ 
       & $0.15^{+0.10}_{-0.06}{}^{+0.11}_{-0.07}$
       &
       &    
       & $0.19^{+0.12}_{-0.07}{}^{+0.09}_{-0.07}$
       & \\[0.2cm]
    $S_-$ 
       &  $1.53^{+0.02}_{-0.02}{}^{+6.55}_{-0.39}$
       & [$0.70^{+0.01}_{-0.01}{}^{+0.43}_{-0.29}$]
       &   
       &  $1.38^{+0.03}_{-0.03}{}^{+0.22}_{-0.17}$
       & [$0.74^{+0.02}_{-0.02}{}^{+0.15}_{-0.11}$] \\
    $S^\prime_- $
       &  $0.51^{+0.10}_{-0.09}{}^{+4.96}_{-0.20}$
       & [$1.19^{+0.16}_{-0.14}{}^{+3.67}_{-0.17}$]
       &    
       &  $0.57^{+0.08}_{-0.07}{}^{+0.10}_{-0.09}$
       & [$1.13^{+0.12}_{-0.10}{}^{+0.09}_{-0.07}$] \\
     $S^{''}_-$ 
       &  $2.98^{+0.56}_{-0.46}{}^{+3.06}_{-2.09}$
       & [$0.59^{+0.07}_{-0.06}{}^{+0.28}_{-0.51}$]
       &    
       &  $2.43^{+0.31}_{-0.28}{}^{+0.56}_{-0.42}$
       & [$0.65^{+0.05}_{-0.05}{}^{+0.09}_{-0.07}$] \\
     \bottomrule
  \end{tabular}
  \let\arraystretch=\oldarraystretch
  \caption{Longitudinal and negative-helicity CP-averaged decay 
    rate ratios as defined in (\ref{sdef}). Numbers in square brackets 
    refer to the unrealistic scenario when the electromagnetic 
    dipole effect is switched off.
  }
  \label{tab:rhokst2}
\end{table}

We define the helicity-specific
CP-averaged decay rate ratios~\cite{BRYEWP}
\begin{equation}
     S_h \equiv \frac{2 \bar \Gamma_h(\rho^0\bar \K^{\ast 0})}
     {\bar \Gamma_h(\rho^-\bar \K^{\ast 0})}, \qquad 
     S_h^\prime  \equiv \frac{2 \bar\Gamma_h(\rho^0\K^{\ast -})}
    {\bar \Gamma_h(\rho^-\K^{\ast 0})},
  \label{sdef}
\end{equation}
and $S_h^{''}\equiv S_h/S_h^\prime$. It should be emphasised that 
the CP-average of helicity-specific decay rates is not the same as the
CP-average of polarisation fractions $f_h$. When the standard
variables are used, the relation involves CP asymmetries. The
$S$-observables defined above are better suited to an investigation of
helicity-specific effects. Experimentally they can be determined from
the same data as the standard observables, thus avoiding unfolding
complicated correlations in the errors of CP asymmetries, branching
and polarisation fractions.  In Table~\ref{tab:rhokst2} we summarise
our theoretical predictions for these ratios, providing results for
both cases (default input and $P_h$ from data), and with the
electromagnetic dipole effect switched off for comparison.  Note that
all ratios would equal 1, if the QCD penguin amplitude was really
dominant. The largely different numbers illustrate the impact of
electroweak penguins in these decays, both for the longitudinal and
transverse amplitudes. One also observes that the electromagnetic
dipole operator contribution is essential in the transverse case.
Using data to fit the leading QCD penguin amplitude from the pure
penguin mode is obviously crucial to discriminate the effect. It is
most pronounced in the $S^{''}_-$ ratio, where the electroweak penguin
amplitude enters the relevant amplitudes with opposite sign, see
(\ref{eq:rhokst}).

\section{\boldmath Other $B\to VV$ decays}
\label{sec:others}

In this section we briefly discuss the remaining 11 decay modes
consisting of electroweak or QCD flavour-singlet penguin-dominated
decays and decays that can only proceed via weak annihilation.  None
of these modes depends on the QCD penguin amplitude $\hat \alpha_4^p$,
so we give the numerical results according to the general procedure
defined in Section~\ref{sec:inputs} and used for the analysis of the
tree-dominated modes in Section~\ref{sec:tree}.

\subsection{\boldmath Other penguin-dominated decays}

These five decays modes are characterised by an interplay of the 
flavour-singlet QCD penguin amplitude $\alpha_3^c$, 
the colour-allowed electroweak penguin amplitude $\alpha_{3,\rm EW}^c$, 
and the colour- and CKM-suppressed tree amplitude 
$\alpha_2$. All three amplitudes are significantly smaller 
than $\alpha_4^c$, hence these decays have small 
branching fractions compared to the $\Delta S=1$ penguin-dominated 
decays.

We first consider the three $\Delta D=1$ 
$\Bm$, $\Bdq$ decay modes in this class, whose 
amplitudes, following the notation of the appendix of \cite{BN2003} and 
neglecting the helicity labels, 
are given by
\begin{eqnarray}
{\cal A}_{B^-\to\rho^-\phi}
  &=& A_{\rho\phi} \Big[ 
      \alpha_3^p - \frac{1}{2}\alpha_{3,{\rm EW}}^p \Big]  , 
\nonumber\\
   -\sqrt{2}\,{\cal A}_{\bar B^0\to\rho^0\phi}
   &=& A_{\rho\phi} \Big[
    \alpha_3^p - \frac{1}{2}\alpha_{3,{\rm EW}}^p\Big], 
\nonumber\\
\sqrt2 \,{\cal A}_{\bar B^0\to\omega\phi}
  &=& A_{\omega\phi} \Big[
   \alpha_3^p - \frac{1}{2}\alpha_{3,{\rm EW}}^p\Big]. 
\end{eqnarray}
The three decay modes have an identical amplitude structure. 
It can be seen from Table~\ref{tab:ampratios} that the QCD penguin 
and electroweak penguin contributions are of similar magnitude for the 
dominant longitudinal amplitude, 
and interfer constructively. Nevertheless, the branching fraction 
of the first decay is expected to be only $(1-3)\cdot 10^{-8}$, 
and the second and third are a factor of two and three smaller, 
respectively. The negative helicity amplitude is dominated by the 
electromagnetic penguin contribution due to the power-enhanced 
effect discussed in \cite{BRYEWP} and Section~\ref{subsec:violation}, 
but it is not larger than the longitudinal amplitude. The 
longitudinal polarisation fraction $f_L$ is predicted in the range 
$0.7\ldots 1$. All CP-violating observables are small, since 
QCD factorisation calculations do not produce a significant 
strong phase between the two terms with different weak phases, 
$\alpha_3^c - \frac{1}{2}\alpha_{3,{\rm EW}}^c$ 
and  $\alpha_3^u - \frac{1}{2}\alpha_{3,{\rm EW}}^u$.
Our results for the branching fractions are compatible with those 
of \cite{Lu:2006si}.

The two $\Bsq$ modes in this class have larger branching fractions, 
since they are governed by $\Delta S=1$ transitions. The amplitudes 
are given by (helicity labels omitted)
\begin{eqnarray}
\sqrt2 \,{\cal A}_{\bar B_s\to \rho^0\phi}
   &=&A_{\phi \rho} \Big[\delta_{pu}\,\alpha_2 + 
    \frac{3}{2}\alpha_{3,{\rm EW}}^p \Big],
\nonumber \\ 
 \sqrt2 \,{\cal A}_{\bar B_s\to\omega\phi}
  &=&A_{\phi\omega} \Big[ 
    \delta_{pu}\,\alpha_2 
    + 2\alpha_3^p + \frac{1}{2}\alpha_{3,{\rm EW}}^p\Big]. 
\end{eqnarray}
Due to a partial cancellation between the QCD and electroweak 
penguin contributions, the CKM-suppressed tree amplitude 
$\alpha_2$ is the largest partial amplitude in the second 
decay. The roles of tree and penguin amplitudes are reversed 
in the first decay. Our results for the various observables 
are summarised in Table~\ref{tab:otherpenguin}. They are 
rather uncertain for $\bar B_s\to\omega\phi$, where the 
non-factorisation of transverse spectator-scattering 
is important, such that often no useful prediction can be 
obtained. It is also worth noting here that we 
have assumed ideal mixing throughout this paper, such that 
the $\omega$ meson has no $s\bar s$ component. Since the 
amplitude for $\bar B_s\to\phi\phi$ is an order of magnitude 
larger than for $\bar B_s\to\omega\phi$ even a small 
mixing angle of about $5^\circ$ could make a significant 
difference in the results for $\bar B_s\to\omega\phi$. The polarisation 
observables of $\bar B_s\to\rho^0\phi$ are determined by 
the power-enhanced contributions from the electromagnetic 
dipole operator, which dominates the transverse electroweak penguin 
amplitude. Similar to $B\to\rho K^*$ decays  \cite{BRYEWP}, 
this contribution changes the sign of $\alpha_{3,\rm EW}^{p-}$ 
relative to naive factorisation, and hence changes $\phi_\parallel$ 
by almost $180^\circ$. 

\begin{table}
  \centering
  \let\oldarraystretch=\arraystretch
  \renewcommand*{\arraystretch}{1.2}
  \begin{tabular}{llll}
    \toprule
        & $\BrAv / 10^{-6}$
        & $f_L$ / percent
        & $\phi_\parallel$ / degrees
       \\
     \midrule
    $\Bsq\to\pphi \prho^0$
        & $0.40^{+0.12}_{-0.10}{}^{+0.25}_{-0.04}$
        & $81^{+3}_{-4}{}^{+9}_{-12}$
        & $\phantom{-}177^{+6}_{-7}{}^{+9}_{-15}$
        \\
    $\Bsq\to\pomega \pphi$
        & $0.10^{+0.05}_{-0.03}{}^{+0.48}_{-0.12}$
        & no prediction
        & $-131^{+1}_{-1}{}^{+51}_{-100}$
        \\[0.2cm]
    \toprule
        & $A_{\rm CP}$ / percent
        & $A_{\rm CP}^0$ / percent
        & $\Delta\phi_\parallel$ / degrees
       \\
     \midrule
    $\Bsq\to\pphi \prho^0$
        & $19^{+5}_{-5}{}^{+56}_{-67}$
        & $11^{+4}_{-3}{}^{+10}_{-8}$
        & $-15^{+4}_{-4}{}^{+19}_{-14}$
        \\
     $\Bsq\to\pomega \pphi$
        & $\phantom{1}8^{+3}_{-3}{}^{+102}_{-56}$
        & no prediction
        &  $\phantom{-}13^{+5}_{-4}{}^{+80}_{-72}$
        \\
   \bottomrule
  \end{tabular}
  \let\arraystretch=\oldarraystretch
  \caption{Predictions for $\Bsq\to\pomega \pphi$ and 
  $\Bsq\to\pphi \prho^0$.}
  \label{tab:otherpenguin}
\end{table}

\subsection{\boldmath Pure weak annihilation decays}

Branching fraction estimates for the six decay modes that can proceed 
only via weak annihilation are given in 
Table~\ref{tab:pureannihilation}. Since QCD factorisation does not 
provide a solid prediction of the annihilation amplitudes, these 
numbers should be regarded as estimates within the 
adopted annihilation model. Measurements of these decay modes 
would result in useful checks of this model. We do not present 
other observables for the annihilation modes, because the 
calculations are too crude to provide quantitative results. 
The following qualitative conclusions can, however, be drawn: 
the decay amplitudes of these modes depend only on the tree 
annihilation amplitude $\beta_1^p$, and the penguin annihilation 
amplitude $\beta_4^p$ (and a corresponding electroweak penguin 
annihilation amplitude), but not on $\beta_3^p$. The calculation 
of the transverse annihilation amplitudes shows an enhancement 
only for $\beta_3^{p-}$ as discussed in Section~\ref{sec:penweakannh}, while 
all others respect the hierarchy 
(\ref{eq:hierarchy}) of helicity amplitudes. Thus we expect 
$f_L$ to be in the range $0.8 \ldots 1$ and $f_\perp\approx f_\parallel$ 
for all modes listed in Table~\ref{tab:pureannihilation}. 
Experimental tests of these expectations would be very interesting 
for the understanding of annihilation dynamics, 
but require rather large data samples.

\begin{table}
  \centering
  \let\oldarraystretch=\arraystretch
  \renewcommand*{\arraystretch}{1.2}
  \begin{tabular}{ll}
    \toprule
        & $\BrAv / 10^{-6}$
    \\
    \midrule
     $\Bdq\to\K^{*-} \K^{*+}$
        & $0.09^{+0.05}_{-0.03}{}^{+0.12}_{-0.10}$
        \\
    $\Bdq\to\pphi \pphi$
        & $<0.03$
        \\\addlinespace
    $\Bsq\to\prho^- \prho^+$
        & $0.34^{+0.03}_{-0.03}{}^{+0.60}_{-0.38}$
        \\
    $\Bsq\to\prho^0 \prho^0$
        & $0.17^{+0.01}_{-0.01}{}^{+0.30}_{-0.19}$
        \\
    $\Bsq\to\pomega \prho^0$
        & $<0.01$
        \\
    $\Bsq\to\pomega \pomega$
        & $0.11^{+0.01}_{-0.01}{}^{+0.20}_{-0.12}$
        \\
    \bottomrule
  \end{tabular}
  \let\arraystretch=\oldarraystretch
  \caption{Branching fraction estimates for pure weak-annihilation 
   modes.}
  \label{tab:pureannihilation}
\end{table}

\section{Conclusion}
\label{sec:conclusion}

In this paper we performed a comprehensive analysis of the 34 $B$
decays to two vector mesons (and their CP conjugates).  Together with
\cite{BN2003}, where the $PP$ and $PV$ final states were discussed,
this completes the phenomenology of two-body decays in QCD
factorisation in next-to-leading order. In comparison with the $PP$
and $PV$ final states, the $VV$ ones are much more uncertain.  This is
due to a potentially large negative-helicity penguin weak-annihilation
amplitude pointed out in \cite{Kagan:2004uw}, but also due to the
non-factorisation of spectator-scattering for the transverse
amplitudes, which has a particularly large effect on colour-suppressed
partial amplitudes. Our main results are summarised as follows.

\subsubsection*{\it Results related to tree-dominated decays}

We obtain a very good description of the $\rho\rho$ system 
including the $\rho^0\rho^0$ final state. As a general rule, 
all colour-suppressed tree decays are poorly predicted due to 
the non-factorisation of transverse spectator-scattering. The 
observed large longitudinal polarisation $f_L(\rho^0\rho^0)$ 
suggests, however, that this effect is not as large as it 
could be. In contrast, the colour-allowed tree-dominated decays are the 
  theoretically best predicted $B\to VV$ modes, and should 
all show $f_L$ near 1. We also find a small longitudinal QCD 
  penguin amplitude, which makes the time-dependent 
  CP asymmetry $S_{L}^{\rho\rho}$ an ideal observable 
to determine the CKM angle $\gamma$. We find
\begin{equation}
  \gamma = (73.2^{+7.6}_{-7.7})^\circ
\qquad \mbox{or} \qquad 
  \alpha = (85.6^{+7.4}_{-7.3})^\circ,
\end{equation}
where the theoretical error alone is only $\pm 3^\circ$.

\subsubsection*{\it Results related to penguin-dominated decays}

The penguin-dominated decays are plagued by the weak-annihilation
uncertainty. While it is natural to obtain an equal amount of
transverse and longitudinal polarisation, many observables can only be
predicted if at least some information is taken from data.  In our
analysis, we explored the possibility of replacing the calculated
negative-helicity penguin amplitude by a fit from $\phi K^*$ data.
While weak annihilation remains the most plausible dynamical
explanation for significant transverse polarisation, other options
exist: spectator-scattering may enhance the transverse flavour-singlet
QCD penguin amplitude, and the electroweak penguin amplitude receives
a large contribution from the electromagnetic dipole
operator~\cite{BRYEWP}. The latter effect is expected to be most
prominent in the $\rho K^*$ system, and we propose to measure certain
helicity-specific decay rate ratios to isolate it.

\vskip0.5cm\noindent
Finally, we comment on the possibility to uncover the helicity 
structure of the weak interactions through polarisation studies 
  in $B\to VV$ decays. For instance, the presence of tensor 
operators changes the helicity-amplitude 
hierarchy (\ref{eq:hierarchy}). According to the theoretical 
  picture that emerges from our study, this appears to be 
very difficult as far as the hierarchy between the negative-helicity 
  and longitudinal amplitude is concerned, since it 
is already weak or violated in the 
QCD penguin amplitudes due to Standard Model QCD dynamics. 
  The tree amplitudes cannot receive large new contributions, 
since similar effects should then be seen in semi-leptonic 
decays. This leaves the colour-allowed electroweak penguin 
  amplitude, which is theoretically well-controlled, providing 
further motivation for the investigation of the $\rho K^*$ 
system. Regarding the hierarchy between the negative- 
  and positive-helcity amplitudes, there is currently no 
  indication of its violation, neither from the $\phi K^*$ 
  data nor from theory. However, one should be aware that 
the factorisation properties of the positive-helicity amplitude 
  are virtually unknown.

\subsubsection*{Acknowledgements}

This work is supported by the Deutsche Forschungsgemeinschaft, SFB/TR~9
``Computer\-ge\-st\"utzte Theo\-re\-ti\-sche Teilchenphysik'', and by the
German-Israeli Foundation for Scientific Research and Development
under Grant No. I - 781-55.14/2003.  D.Y. acknowledges support from
the Japan Society for the Promotion of Science.

\subsubsection*{Note added}

After publication of the preprint version of this paper, we have been 
made aware of talks by A.~Kagan, where a fit to the $B\to\phi K^*$
system similar to the one discussed in Section~\ref{phikstsec} was 
reported, arriving at similar conlcusions on the magnitude of 
transverse weak annihilation.


\appendix

\section{Appendix}

In this appendix we collect the formulae for the hard-scattering functions 
relevant to the negative-helicity amplitudes.

\subsection{Light-cone projection}

Up to twist-3, six two-particle light-cone distribution amplitudes are
relevant for vector mesons. We neglect three-particle $q\bar q g$
amplitudes. Consistency then requires that one adopts the 
Wandzura-Wilczek relations among the two-particle amplitudes, reducing 
the input to two amplitudes $\Phi_V\equiv \phi_\parallel$ and 
$\phi_\perp$ \cite{Ball:1998sk}.

The two-particle light-cone projection operator on the longitudinal 
polarisation state of the vector meson is given as 
$M^V_\parallel$ given in \cite{BN2003}, Section 2.3. To obtain 
the projector on the transverse polarisation states in the 
helicity basis, we use the result for $M^V_\perp$ 
from \cite{BF2000}, and insert $\epsilon_\perp=\epsilon_\mp$ to 
obtain 
\begin{equation}
 \begin{split}
  M_\mp^V(u) = 
    &-\frac{i f_\perp}{4}\FMslash{\epsilon}^*_\mp\!\FMslash{p} 
     \;\phi_\perp(u)\\
    &-\frac{i f_V m_V}{8} \Bigg\{\!
     \FMslash{\epsilon}^*_\mp (1-\gamma_5)
     \left(g_\perp^{(v)}(u)\pm\frac{g^{(a)\prime}_\perp}{4}\right)
     + \FMslash{\epsilon}^*_\mp  (1+\gamma_5)
     \left(g_\perp^{(v)}(u)\mp\frac{g^{(a)\prime}_\perp}{4}\right) \\
    &\phantom{-\frac{i f_V m_V}{4} \Bigg\{\;}
        -\FMslash{p} \,(1-\gamma_5) \left(\int_0^u \d v \,(\Phi_V(v) - 
        g_\perp^{(v)}(v)) \mp \frac{g_\perp^{(a)}(u)}{4}
         \right) \epsilon^{*}_{\mp\nu} 
        \frac{\partial}{\partial l_{\perp\nu}}    \\
    &\phantom{-\frac{i f_V m_V}{4} \Bigg\{\;}
        -\FMslash{p} \,(1+\gamma_5) \left(\int_0^u \d v \,(\Phi_V(v) - 
        g_\perp^{(v)}(v)) \pm \frac{g_\perp^{(a)}(u)}{4}
         \right) \epsilon^{*}_{\mp\nu} 
        \frac{\partial}{\partial l_{\perp\nu}}
      \Bigg\}\Bigg|_{l_\perp=0},
 \end{split}
\label{longprojector}
\end{equation}
where $p$ is the momentum of the vector meson, and 
we assigned the parton momenta
\begin{equation}
  \label{eq:quarkmomenta}
       l  = u p + l_\perp - \frac{l_\perp^2}{u m_B^2}\bar{p}
  \qquad\text{and}\qquad
  \bar{l} = \bar{u} p - l_\perp - \frac{l_\perp^2}{\bar{u} m_B^2}\bar{p}
\end{equation}
to the quark and antiquark constituents, respectively. 
The collinear approximation $l_\perp=0$ must be applied only 
after the projection. Using the Wandzura-Wilczek relations 
\cite{Ball:1998sk} and defining \cite{Kagan:2004uw}
\begin{equation}
  \phi_a(u) = \int_u^1 \d{v}\; \frac{\Phi_V(v)}{v},
  \qquad\quad
  \phi_b(u) = \int_0^u \d{v}\; \frac{\Phi_V(v)}{\bar{v}},
  \label{eq:defphiabv}
\end{equation}
the transverse helicity projectors simplify to
\begin{equation}
 \label{eq:Mperp}
 \begin{split}
  M_\mp^V(u) = 
    &-\frac{i f_\perp}{4}\FMslash{\epsilon}^*_\mp\!\FMslash{p} 
     \;\phi_\perp(u)\\
    &-\frac{i f_V m_V}{8} \Bigg\{
     \epsilon^*_{\mp\nu} \,\phi_a(u) \left[\gamma^\nu (1\mp\gamma_5) 
    +u\!\FMslash{p} \,(1\mp \gamma_5)\frac{\partial}{\partial l_{\perp\nu}}
     \right]\\
    &\hspace*{2cm} 
    + \epsilon^*_{\mp\nu} \,\phi_b(u) \left[\gamma^\nu (1\pm\gamma_5) 
    -\bar u\!\FMslash{p} \,(1\pm \gamma_5)
    \frac{\partial}{\partial l_{\perp\nu}}\right]
 \Bigg\}\Bigg|_{l_\perp=0}.
 \end{split}
\end{equation}
Many previous calculations of transverse polarisation amplitudes
\cite{Cheng:2001aa,Li:2003he,Yang:2004pm,Zou:2005gw}  
used an incorrect expression for the projector, where the 
transverse-derivative terms are dropped. The authors of 
\cite{Das:2004hq} used (\ref{longprojector}) without employing 
the Wandzura-Wilczek relations. The projector is not given 
explicitly in \cite{Kagan:2004uw}.

\boldmath
\subsection{The $a_i^{p,h}$ coefficients}
\unboldmath

The leading order results for the coefficients $a_i^{p,h}$, needed to
calculate amplitudes according to (\ref{ampexample}),
(\ref{eq:alphaa}), reproduce the results from naive factorisation. At
next-to-leading order they receive contributions from one-loop vertex
corrections, penguin and dipole operator insertion topologies and hard
spectator interaction terms.  We assemble these in the form
\begin{equation}
  \begin{split}
    a_i^{p,h}(V_1V_2)
        &=  \left(C_i + \frac{C_{i\pm 1}}{\Nc}\right) N^h_i(V_2)
            \\&\mathrel{\hphantom{=}}
           +\frac{C_{i\pm 1}}{\Nc} \frac{\CF\alphas}{4\pi}
            \left[V_i^h(V_2) + \frac{4\pi^2}{\Nc}H_i^h(V_1 V_2)
            \right]
           + P^{h,p}_i(V_2),
  \end{split}
\label{ai}
\end{equation}
where the upper (lower) signs apply when $i$ is odd (even), and 
it is understood that the superscript `p' is to be ommited for
$i\in\{1,2\}$.

The results for $h=0$ correspond to those given in~\cite{BN2003} for
$PV$ final states with obvious replacements of $P$ by $V$, so we will 
only give explicit results for the negative-helicity amplitude. 
The leading-order coefficient, corresponding to naive factorisation, 
is simply
\begin{equation}
   N_i^- (V_2) = 
     \begin{cases}
       0, & i\in\{6,8\}\\
       1, & \text{else.}
     \end{cases}
\end{equation}
As is well-known, there is no leading-order 
contribution from $(S-P)\otimes(S+P)$ operators for vector mesons. 

The negative-helicity vertex corrections read
\begin{equation}
  \label{eq:V}
  V^-_i(V_2) = \left\{
    \begin{array}{ll}
       \displaystyle
       \int_0^1 \d y\; \phi_{b2}(y) 
          \left[ \hphantom{-} 
                 12\ln\frac{m_b}{\mu} - 18 + g_T(y) \right],
       &\qquad i \in \{1,2,3,4,9,10\} 
       \\[1em]
       \displaystyle
       \int_0^1 \d y\; \phi_{a2}(y) 
          \left[ -12\ln\frac{m_b}{\mu} + \hphantom{0}6 - g_T(\bar{y}) \right],
       &\qquad i \in \{5,7\}
       \\[1.7em]
       \displaystyle
       0, &\qquad i \in \{6,8\}
    \end{array}
  \right.
\end{equation}
with ($\bar y\equiv 1-y$; similarly for other convolution variables 
below)
\begin{eqnarray}
    g_T(y) &= &\frac{4-6y}{\bar{y}}\ln y - 3 i \pi 
              + \left(2\Li_2(y) - \ln^2 y + \frac{2\ln y}{\bar{y}}
                    - (3+2\pi i) \ln y
                    - [ y\to\bar{y}]
                 \right) \nonumber \\ 
          &=& g(y)+\frac{\ln y}{\bar y}. 
\end{eqnarray}
The function $g_T(y)$ differs from the corresponding function $g(y)$ 
in the longitudinal amplitude only by a single term. 

Corrections from penguin contractions and dipole-operator insertions
are present for $i\in\{4,\text{7--10}\}$ at order $\alphas$
($\alphaEM$ for the electroweak penguin amplitudes $i=\text{7--10}$).
We find $ P_6^{-,p}(V_2)= P_8^{-,p}(V_2)=0$, and
\begin{align}
  \label{eq:P}
  \begin{split}
    P_4^{-,p}(V_2) &= \frac{\alphas \CF}{4\pi \Nc} 
      \bigg\{
         C_1 \left[\frac{2}{3}\ln\frac{m_b^2}{\mu^2} + \frac{2}{3}
                   - G^{-}_{V_2}(s_p)
             \right]\\&\hphantom{=\frac{\alphas \CF}{4\pi \Nc}\bigg\{}
        +C_3 \left[\frac{4}{3}\ln\frac{m_b^2}{\mu^2} + \frac{4}{3}
                   - G^{-}_{V_2}(0) - G^{-}_{V_2}(1)
             \right]\\&\hphantom{=\frac{\alphas \CF}{4\pi \Nc}\bigg\{}
        +(C_4+C_6) \left[\frac{10}{3}\ln\frac{m_b^2}{\mu^2} 
                   - 3 G^{-}_{V_2}(0) - G^{-}_{V_2}(s_c) 
                   - G^{-}_{V_2}(1)
             \right]
      \bigg\},
  \end{split}\\[0.3cm]
    P_{7}^{-,p}(V_2) &= P_{9}^{-,p}(V_2) =  -\frac{\alphaEM}{3\pi} 
       C^{\mathrm{eff}}_{7\gamma}\,
       \frac{m_B m_b}{m_2^2} 
     \nonumber\\
     &\hspace*{2.4cm}+\frac{2 \alphaEM}{27\pi}\,(C_1+\Nc C_2)
       \left[\delta_{pc} \ln\frac{m_c^2}{\mu^2} + 
             \delta_{pu} \ln\frac{\nu^2}{\mu^2} + 1\right],\\[0.3cm]
    P_{10}^{-,p}(V_2) &= \frac{\alphaEM}{9\pi \Nc}
     \bigg\{(C_1+\Nc C_2)\left[
                       \frac{2}{3}\ln\frac{m_b^2}{\mu^2} + \frac{2}{3}
                       -G^{-}_{V_2}(s_p)
                     \right]
     \bigg\}
\end{align}
with
\begin{equation}
 \label{eq:penguinconvolutions}
  G^-_{V_2}(s) = \int_0^1 \d{y}\; \phi_{b2}(y) G(s- i\epsilon,\bar{y}),
\end{equation}
where $G(s,y)$ is the usual penguin function (see, for instance,
\cite{BBNS2001}). Notice that unlike the longitudinal case, there is
no contribution from the dipole operators $Q_{8g}$ and $Q_{7\gamma}$
to $ P_{4}^{-,p}(V_2)$, $P_{10}^{-,p}(V_2)$, for which we confirm the
result given in~\cite{Kagan:2004uw}. The two terms $
P_{7}^{-,p}(V_2)$, $P_{9}^{-,p}(V_2)$ exhibit an enhancement by
$(m_b/\LQCD)^2$ proportional to $C^{\mathrm{eff}}_{7\gamma}$ due to
the factor $m_B m_b/m_2^2$, which alters the naive power-counting
(\ref{eq:hierarchy}) to (\ref{eq:emhierarchy}).  See also
\cite{BRYEWP}. We neglect the small contributions from the electroweak
penguin operators $Q_{7-10}$ to the penguin coefficients.

The spectator-scattering contribution is given by
\begin{align}
  \label{eq:Hm1}
  H^-_{i} &=
      - \frac{2 f_B f_{V_1}^\perp}{m_B m_b F_-^{\B\to V_1}(0)}\,
      \frac{m_b}{\lambda_B}
      \int_0^1 \d{x} \d{y} \;
           \frac{\phi_1^\perp(x) \phi_{b2}(y)}{\bar{x}^2 y} \\
  \intertext{for $i\in \{1,2,3,4,9,10\}$,}
  \label{eq:Hm2}
  H^-_{i} &=
      + \frac{2 f_B f_{V_1}^\perp}{m_B m_b F_-^{\B\to V_1}(0)}\,
      \frac{m_b}{\lambda_B}
      \int_0^1 \d{x} \d{y} \;
           \frac{\phi_1^\perp(x) \phi_{a2}(y)}{\bar{x}^2 \bar{y}}
  \\
  \intertext{for $i\in\{5,7\}$ and }
  \label{eq:Hm3}
  H^-_{i} &= 
      + \frac{f_B f_{V_1}}{m_B m_b F_-^{\B\to V_1}(0)}\,
      \frac{m_b m_1}{m_2^2}\,\frac{m_b}{\lambda_B}
      \int_0^1 \d{x} \d{y} \;
           \frac{\phi_{a1}(x) \phi_2^\perp(y)}{y \bar{x} \bar{y}}
\end{align}
for $i\in\{6,8\}$. Our expressions for the spectator-scattering
kernels are simpler than previously published results
\cite{Cheng:2001aa,Li:2003he,Yang:2004pm,Das:2004hq,Zou:2005gw}, and
differ even when the Wandzura-Wilzcek relations are used to simplify
those results. The $\B$ meson light-cone distribution amplitude enters
the kernels via the parameter $\lambda_B$ defined in \cite{BBNS1999}.
The factors in front of the integrals are of order 1 in the
$\LQCD/m_b$ counting except for $i=6,8$, where there is an extra
$m_B/\LQCD$-enhancement.  Thus, the hard-spectator scattering
contribution to $a_{6,8}^{p-}$ is formally leading over the
naive-factorisation amplitude, although not numerically, because
$H_{6,8}^-$ are multiplied by small Wilson coefficients. Since
$a_{6,8}^{p-}$ contribute to a decay amplitude in the product
$r_\chi^{V_2} a_{6,8}^{p-}$, and since $r_\chi^{V_2}\sim \LQCD/m_b$
according to (\ref{rchidef}), we conclude that the suppression of the
(pseudo)scalar penguin amplitudes relative to the $V-A$ ones,
$a_4^{p-}$, is absent for the negative helicity amplitude.

Another qualitative difference to the longitudinal amplitude is that
the $x$-integrals in (\ref{eq:Hm1}), (\ref{eq:Hm2}) are divergent,
since the integrand is too singular near $x=1$.  Thus factorisation
for the negative-helicity amplitude breaks down even at leading order
in the heavy-quark expansion due to the non-factorisation of
spectator-scattering. To estimate this contribution, we extract the
logarithmic divergence by applying a ``plus-prescription'' to the
integrand, and replace the large logarithm by a phenomenological
parameter $X_H^{V_1}$ \cite{BN2003}. To estimate the endpoint
behaviour, we note that the asymptotic distribution amplitudes are
$\phi^\perp(x) \to 6 x\bar x$, $\phi_a(x)\to 3\bar x^2$, $\phi_b(x)\to
3 x^2$. We can then write
\begin{equation}
    \int_0^1 \d{x}\; \frac{\phi^\perp_1(x)}{\bar{x}^2} =
      \left(\lim_{u\to 1} \frac{\phi^\perp_1(u)}{\bar{u}}\right)  X_H^{V_1}
    + \underbrace{\int_0^1 \frac{\d{x}}{\bar{x}}\left[
                     \frac{\phi^\perp_1(x)}{\bar{x}}
    -\left(\lim_{u\to 1} \frac{\phi^\perp_1(u)}{\bar{u}}\right)
                  \right]}%
      _{\text{finite}}.
\end{equation}
As specified in (\ref{XHparam}), we use a simple model where we treat
$X_H$ as an unknown complex parameter universal to all $H_i(V_1 V_2)$
with magnitude around $\ln(m_B/\LQCD)$ because the logarithmic
infrared divergence has its origin in a soft gluon interaction with
the spectator quark and can therefore be expected to be regulated at a
physical scale of order \LQCD.

\boldmath
\subsection{Weak annihilation contributions ($b_i^{p,h}$ coefficients)}
\unboldmath

Weak-annihilation is parameterised by a set of amplitudes $b_i^p(V_1
V_2)$. The leading contributions can be assembled from a few basic
building blocks as shown in~\cite{BN2003,BBNS2001}. The corresponding
formulae also hold for the helicity amplitudes in $B\to VV$ decays,
hence we only summarise these building blocks here.

For the longitudinal case $h=0$, only a few
signs change in comparison with the known results for $\B\to PP$ or
$PV$. We find 
\begin{equation}
  \begin{split}
    \Ai[,0]_1 &= \pi\alphas\int_0^1\d{x}\d{y}\left\{
                    \Phi_{V_1}(x) \Phi_{V_2}(y) \left[
                       \frac{1}{x(1-\bar{x}y)}+\frac{1}{x\bar{y}^2}
                    \right]
                    -r_\chi^{V_1} r_\chi^{V_2} \Phi_{v1}(x)\Phi_{v2}(y)
                       \frac{2}{x\bar{y}}
                 \right\},\\[0.4cm]
    \Ai[,0]_2 &= \pi\alphas\int_0^1\d{x}\d{y}\left\{
                    \Phi_{V_1}(x) \Phi_{V_2}(y) \left[
                       \frac{1}{\bar{y}(1-\bar{x}y)}+\frac{1}{x^2\bar{y}}
                    \right]
                    -r_\chi^{V_1} r_\chi^{V_2} \Phi_{v1}(x)\Phi_{v2}(y)
                       \frac{2}{x\bar{y}}
                 \right\},\\[0.4cm]
    \Ai[,0]_3 &= \pi\alphas\int_0^1\d{x}\d{y}\left\{
                     r_\chi^{V_1} \Phi_{v1}(x) \Phi_{V_2}(y)
                        \frac{2\bar{x}}{x\bar{y}(1-\bar{x}y)}
                   + r_\chi^{V_2} \Phi_{V_1}(x) \Phi_{v2}(y)
                        \frac{2y}{x\bar{y}(1-\bar{x}y)}
                 \right\},\\[0.4cm]
    \Af[,0]_3 &= \pi\alphas\int_0^1\d{x}\d{y}\left\{
                     r_\chi^{V_1} \Phi_{v1}(x) \Phi_{V_2}(y)
                        \frac{2(1+\bar{y})}{x\bar{y}^2}
                   - r_\chi^{V_2} \Phi_{V_1}(x) \Phi_{v2}(y)
                        \frac{2(1+x)}{x^2\bar{y}}
                 \right\},
  \end{split}
\end{equation}
and $\Af[,0]_1 = \Af[,0]_2 = 0$. The non-vanishing transverse building
blocks are
\begin{equation}
  \begin{split}
    \Ai[-]_1 &= \pi \alphas\, \frac{2 m_1 m_2}{m_B^2} \int_0^1 \d{x}\d{y} \;
       \Bigg\{ 
           \phi_{b1}(x) \phi_{b2}(y)  
           \left[  \frac{\bar{x}+\bar{y}}{x^2\bar{y}^2} + 
                   \frac{1}{(1-\bar{x}y)^2}
           \right]
       \Bigg\},\\[0.4cm]
    \Ai[-]_2 &= \pi \alphas\, \frac{2m_1 m_2}{m_B^2} \int_0^1 \d{x}\d{y} \;
       \left\{ \phi_{a1}(x)\phi_{a2}(y)
            \left[ \frac{x+y}{x^2\bar{y}^2}
                   + \frac{1}{(1-\bar{x}y)^2}
            \right]
       \right\},\\[0.4cm]
    \Ai[-]_3 &=  \pi \alphas \int_0^1 \d{x}\d{y} \;
        \left\{
           \frac{2m_1}{m_2}  r_\chi^{V_2}
           \frac{  \phi_{a1}(x) \phi_2^\perp(y) }%
                {x \bar{y} (1-\bar{x}y)}
         - \frac{2m_2}{m_1}  r_\chi^{V_1}
           \frac{  \phi_1^\perp(x) \phi_{b2}(y) }%
                {x \bar{y} (1-\bar{x}y)}
       \right\}, \\[0.4cm]
    \Af[-]_3 &=  \pi \alphas \int_0^1 \d{x}\d{y} \;
       \left\{
           \frac{2m_1}{m_2} r_\chi^{V_2} \frac{\phi_{a1}(x)\phi_2^\perp(y)}%
                {x\bar{y}^2}
         + \frac{2m_2}{m_1} r_\chi^{V_1} \frac{\phi_1^\perp(x)\phi_{b2}(y)}%
                {x^2 \bar{y}}
       \right\},\\[0.4cm]
    \Ai[+]_1 &= \pi \alphas\, \frac{2 m_1 m_2}{m_B^2} \int_0^1 \d{x}\d{y} \;
       \Bigg\{ 
           \phi_{a1}(x) \phi_{a2}(y)
           \left[ \frac{2}{x\bar{y}^3}-\frac{y}{(1-\bar{x}y)^2}
                 -\frac{y}{\bar{y}^2(1-\bar{x}y)}
           \right]  
       \Bigg\},\\[0.4cm]
    \Ai[+]_2 &= \pi \alphas\, \frac{2 m_1 m_2}{m_B^2} \int_0^1 \d{x}\d{y} \;
       \left\{ \phi_{b1}(x)\phi_{b2}(y)
            \left[ \frac{2}{x^3\bar{y}} - \frac{\bar{x}}{(1-\bar{x}y)^2}
                  -\frac{\bar{x}}{x^2(1-\bar{x}y)}
            \right]
       \right\}.
  \end{split}
\end{equation}
Again, the convolution integrals
exhibit logarithmic and even linear infrared divergences, which we
extract into unknown complex quantities using the prescriptions
\begin{equation}
  \label{eq:XA}
  \int_0^1 \frac{\d{u}}{u} \to X_A,
  \qquad\qquad
  \int_0^1\d{u}\,\frac{\ln u}{u} \to -\frac{1}{2} \left(X_A\right)^2,
  \qquad\qquad
  \int_0^1 \frac{\d{u}}{u^2} \to X_L.
\end{equation}
As for $X_H$, we assume both $X_A$ and $X_L$ to be universal to
all $VV$ final states with magnitudes around
$\ln(m_b/\LQCD)$ and $m_b/\LQCD$, respectively.

Since all non-vanishing building blocks contain such
divergences, making the treatment of annihilation rather model dependent, 
we further simplify our results by evaluating the
convolution integrals with asymptotic distribution amplitudes
$\Phi_V(u)=\phi_\perp^V(u)= 6u\bar{u}$,
$\phi_a(u)=\phi_b(\bar{u})=3\bar{u}^2$, $\Phi_v(u)=3(u-\bar{u})$ and
obtain the expressions
\begin{equation}
  \label{eq:annihilationuniversal1}
  \begin{split}
    \Ai[,0]_1 \approx \Af[,0]_2 &\approx
         18\pi\alphas\left[
             \left(X_A-4+\frac{\pi^2}{3}\right)
           + r_\chi^{V_1} r_\chi^{V_2} \left(X_A-2\right)^2
         \right],\\
    \Ai[,0]_3 &\approx 18\pi\alphas\left(r_\chi^{V_1}+r_\chi^{V_2}\right)
                       \left(-X_A^2 + 2X_A - 4 + \frac{\pi^2}{3} 
                       \right),\\[0.1cm]
    \Af[,0]_3 &\approx 18\pi\alphas\left(r_\chi^{V_1}-r_\chi^{V_2}\right)
                       \left(2X_A-1\right)\left(2-X_A\right)
  \end{split}
\end{equation}
for the nonvanishing longitudinal contributions, and
\begin{equation}
  \label{eq:annihilationuniversal2}
  \begin{split}
    \Ai[+]_1 \approx \Ai[+]_2 &\approx 
         18\pi\alphas \frac{m_1 m_2}{m_B^2}\left(
             2X_A^2 - 3X_A + 6 - \frac{2}{3}\pi^2 \right), \\[0.1cm]
    \Ai[-]_1 \approx \Ai[-]_2 &\approx 
         18\pi\alphas \frac{m_1 m_2}{m_B^2}\left(
             \frac{1}{2}X_L + \frac{5}{2} - \frac{\pi^2}{3} \right),
       \\[0.1cm]
    \Ai[-]_3 &\approx 18\pi\alphas
          \left(\frac{m_1}{m_2}r_\chi^{V_2}-\frac{m_2}{m_1}
          r_\chi^{V_1}\right)
          \left( X_A^2 - 2X_A + 2 \right),\\[0.1cm]
    \Af[-]_3 &\approx 18\pi\alphas
          \left(\frac{m_1}{m_2}r_\chi^{V_2}+\frac{m_2}{m_1}
          r_\chi^{V_1}\right)
          \left( 2X_A^2 - 5 X_A + 3 \right)
  \end{split}
\end{equation}
for the transverse ones. As expected, all annihilation contributions
are suppressed by at least one power of $\LQCD/m_B$ compared to the
form factor contributions, with $\Ai[,0]_3, \Af[,0]_3$ and
$\Ai[+]_{1,2}$ carrying an additional explicit suppression factor.
Where a comparison is possible, these results agree with
\cite{Kagan:2004uw}.



\end{document}